\documentclass[a4paper,11pt]{article}
\pdfoutput=1 

\usepackage{jcappub} 

\usepackage[T1]{fontenc} 

\usepackage{comment}

\providecommand{\beq}{\begin{equation}}
\providecommand{\eeq}{\end{equation}}
\providecommand{\abs}[1]{\lvert#1\rvert}

\newcommand{\vv}{\mbox{\bf {v}}}

\newcommand{\vnh}{\hat{\mbox{\bf {n}}}}

\newcommand{\gsim}{\; ^{>}_{\sim}\;}

\newcommand\aap{A\&A}                
\newcommand\aj{AJ}                   
\newcommand\ao{Appl. Opt.}           
\newcommand\apj{ApJ}                 
\newcommand\apjl{ApJ}                
\newcommand\apjs{ApJS}               
\newcommand\apss{Ap\&SS}             
\newcommand\mnras{MNRAS}             
\newcommand\prd{Phys. Rev.~D}        

\title{\boldmath Relativistic Angular Redshift Fluctuations embedded in Large Scale Varying Gravitational Potentials.}

\author[a,b]{Adal Lima-Hernández,}
\author[a,b,1]{Carlos Hernández-Monteagudo,\note{Corresponding author.}}
\author[c]{and Jon\'as Chaves-Montero}

\affiliation[a]{Instituto de Astrofísica de Canarias (IAC), \\ C/Vía Láctea, s/n, E-38205, La Laguna, Tenerife, Spain }
\affiliation[b]{Departamento de Astrofísica, Universidad de La Laguna, \\Avenida Francisco S\'anchez, s/n, E-38205, La Laguna, Tenerife, Spain}
\affiliation[c]{Donostia International Physics Center, \\ Paseo Manuel de Lardizábal 4, E-20018 Donostia-San Sebastián, Spain}
\emailAdd{alu0101118307@ull.edu.es}
\emailAdd{chm@iac.es}

\abstract{We compute the linear order, general relativistic corrections to angular redshift fluctuations (ARF), a new cosmological observable built upon density-weighted two-dimensional (2D) maps of galaxy redshifts. We start with an existing approach for galaxy/source counts developed in the Newtonian gauge, and generalize it to ARF, modifying for this purpose a standard Boltzmann code. Our calculations allow us identifying the velocity terms as the leading corrections on large scales, emphasizing the sensitivity of ARF to peculiar, cosmological velocity fields. Just like for standard 2D clustering, the impact of gravitational lensing on ARF is dominant on small angular scales and for wide redshift shells, while the signatures associated to gravitational potentials are extremely small and hardly detectable. The ARF also present interesting correlation properties to anisotropies of the Cosmic Microwave Background (CMB): they are highly correlated to CMB lensing potential fluctuations, while also exhibiting a significant (S/N$\sim 4$--$5$) {\em anti-}correlation with the Integrated Sachs-Wolfe effect (ISW). This negative ARF$\times$ISW signal is quite complementary to the standard 2D clustering$\times$ISW correlation, since the former appears mostly at higher redshift ($z\sim 2$) than the latter ($z\lesssim 1)$, and the combination of the two observables significantly increases the $\chi^2$ statistics testing the null (no ISW) hypothesis. We conclude that ARF constitute a novel, alternative, and potentially powerful tool to constrain the nature of Dark Energy component that gives rise to the ISW.
}

\begin{document}
\maketitle
\flushbottom

\section{Introduction}

Cosmologists study the Universe using different, complementary observational avenues to improve the constraining power of each observable by itself. Sub-millimeter observations of the Cosmic Microwave Background (CMB) from ground- and space-based experiments such as {\it Planck} \cite{planck14a}, ACTPol \cite{niemack2010_ACTPolPolarizationsensitivereceiver}, SPT-3G \cite{benson2014_SPT3GNextgenerationcosmic}, or BICEP3 \cite{ade2022_BICEPKeckXV} have provided an exquisite view of the primeval, quasi-homogeneous, and quasi-isotropic universe at $z\simeq 1\,050$, setting strict constraints on the parameters defining the standard cosmological model and the inflationary epoch immediately following the Big Bang. Furthermore, these observations are sensitive to the low redshift universe as CMB photons interact with matter in their way towards us, giving rise to a plethora of so-called {\it secondary effects}, including gravitational lensing, which consists in the {\it bending} of CMB geodesics by intervening matter \cite{blanchard1987_GravitationalLensingeffect, cole1989_GravitationalLensingfluctuations}, the Sunyaev-Zeldovich effect \cite{sunyaev70, Sunyaev1972, Sunyaev1980}, which is caused by the scattering of CMB photons off free electrons, and the Integrated Sachs-Wolfe (ISW) effect \cite{sachs67}, which is induced by the time-evolution of large-scale gravitational potentials.

On the other hand, cosmologists study the late universe using optical and infrared observations of galaxies and quasars, aiming to build three-dimensional maps of the underlying density field. Using these surveys of the Large Scale Structure (LSS) of the universe, cosmologists extract precise cosmological information from multiple observables, including the spatial/angular distribution of galaxies \cite{eisenstein05}, the distortions of galaxy shapes caused by gravitational lensing \cite{Miraldaescude91}, and the abundance and size function of under-dense regions \cite{krause2013_WeightEmptinessGravitational} and clusters of galaxies \cite{allen2011_CosmologicalParametersObservations}. Extracting cosmology from LSS surveys requires modeling observational systematics and multiple physical ingredients such as how galaxies trace the matter density field, baryonic effects, and nonlinearities induced by gravity, which present a larger impact on LSS studies than on CMB analyses. In fact, there is an significant ongoing effort to constrain, model and/or marginalize over the uncertainties associated to observational systematics \cite{bridle2007_DarkEnergyconstraints, sheldon2017_PracticalWeaklensingShear, ross17, chavesmontero18} and the aforementioned physical phenomena \cite{zheng05, croton2006_ManyLivesactive, somerville2008_SemianalyticModelcoevolution, schaye15, Springel2018, behroozi2019_UniverseMachineCorrelationGalaxy, contreras2021_FlexibleSubhaloabundance, schneider2015_NewMethodquantify, arico2019_ModellingLargescale}.

Ideally, the cosmological constraints obtained from CMB or LSS surveys should be compatible and complementary. In practice, the analysis of data from the latest CMB and LSS cosmological surveys has given rise to tensions \cite{Planck2018VI} in the value of key cosmological quantities such as the Hubble's expansion parameter, the amplitude linear matter perturbations, and the effective amplitude of gravitational lensing. Because of this, a greater focus has been put on issues like precise error computation, confirmation bias, the impact of known and unknown systematics, and consistency tests via alternative cosmological probes. Among the latter, several new observables have been proposed. The abundance, shape and size, and spatial distribution of voids have proven to be highly cosmology-sensitive and have been the subject of research in the last few years \cite{cai2015_TestingGravityusing, hamaus2020_PrecisionCosmologyvoids, hamaus2020_PrecisionCosmologyvoids}. Cosmic chronometers, first introduced by \cite{jimenez2002_ConstrainingCosmologicalParameters}, are also becoming more widely used in cosmological parameter estimation \cite{stern2009_CosmicChronometersConstraining, nunes2016_NewConstraintsinteracting, moresco2018_SettingStageCosmic, moresco2020_SettingStageCosmic}. The use of Fast Radio Bursts (FRBs) \cite{gao2014_FastRadioBurst, munoz2016_LensingFastRadio, li2018_StronglyLensedrepeating, jaroszynski2019_FastRadiobursts, li2019_CosmologyindependentEstimateFraction}, intensity mapping \cite{battye2004_NeutralHydrogensurveys, loeb2008_PossibilityPreciseMeasurement, chang2008_BaryonAcousticOscillation}, and redshift drifts \cite{sandage1962_ChangeRedshiftApparent, loeb1998_DirectMeasurementCosmological} have also been suggested in a cosmological context.

This work focuses on the study of ``density-weighted Angular Redshift Fluctuations" (ARF), a novel cosmological observable first introduced by \cite{hernandez-monteagudo2021_DensityWeightedangular} (Letter I hereafter) that refers to fluctuations in the redshift field sampled by galaxies selected under a particular redshift window. As shown in Letter I, fluctuations around the average redshift of sources selected under a window are sensitive to the growth rate of perturbations, large-scale galaxy bias, and the level of primordial non-Gaussianity. The constraining power of ARF was first shown in \cite{hernandez-monteagudo2021_TomographicConstraintsgravity} (Letter II hereafter), where the authors leveraged the sensitivity of ARF to peculiar radial velocities to set strict constraints on the nature of gravity using galaxies from the Baryon Oscillation
Spectroscopic Survey \cite{dawson13}. The constraining power of ARF for upcoming LSS surveys Dark Energy Spectroscopic Instrument (DESI) \cite{desicollaboration2016_DESIExperimentParta} and Euclid \cite{laureijs11} will be even more significant, and the combination of ARF and galaxy clustering has the potential to deliver ten times more strict constraints on the time-evolution of dark energy than galaxy clustering on its own \cite{legrand2021_HighresolutionTomographygalaxy}. Furthermore, \cite{chaves-montero2021_MeasuringEvolutionintergalactic} used the cross-correlation of ARF maps built upon state-of-the-art spectroscopic surveys and CMB maps from the {\it Planck} experiment to extract the highest significance detection (S/N$\simeq 11$) of the kinetic Sunyaev-Zeldovich \cite{Sunyaev1980} effect up to date and characterize the distribution of intergalactic gas from $z\simeq0.1$ to 5.

In this work, we further investigate the potential of the ARF by studying the linear order corrections from General Relativity (GR) \cite{einstein1916}. Following the same strategy as \cite{challinor2011_LinearPowerspectrum} for density fluctuations, we derive first-order GR terms for ARF. Then, we modify the Boltzmann code {\tt CAMB sources} \cite{2011ascl.soft05013C} to estimate the ARF auto-angular power spectra and the cross-correlation of ARF and primordial CMB anisotropies. Finally, we investigate the potential of using ARF to measure the ISW effect, finding that the cross-correlation of ARF and CMB observations is sensitive to this effect for redshifts at which the sensitivity of the cross-correlation of galaxy clustering and CMB observations is low. These results let us conclude that the combination of ARF and galaxy clustering yields significantly more precise measurements of the ISW effect than galaxy clustering alone.

About a week before the submission of this work, \cite{matthewson2022_RedshiftWeightedgalaxy} have presented an independent computation of the GR linear corrections to ARF: despite we are working on different gauges, the comparison of their different correcting terms to ours show very good agreement (as it will be discussed below). Their forecast work seems also to be in good agreement with that of \cite{legrand2021_HighresolutionTomographygalaxy}, at least for the cases where the parameter configuration in both works match each other.

In Sect.~\ref{sec:PostNewt} we review the post-Newtonian derivation of the ARF, while in Sect.~\ref{sec:ARFinGR} we present the GR description of the ARF in the Newtonian gauge. This section first summarises very briefly the work by \cite{challinor2011_LinearPowerspectrum} (CL11 hereafter), (of which a more detailed description can be found in Appendix~\ref{app:ADF}), and then outlines the basic steps required to obtain the ARF transfer functions from the ADF ones. A more detailed description of this derivation can be found in Appendix~\ref{ap:B}). In Sect.~\ref{sec:results} we present our results for the ARF auto power spectra, and their cross-correlation to CMB observables ($T$, $E$-mode of polarization, and the $\phi$ deflection potential). Finally, in Sect.~\ref{sec:discussion} we discuss our results in the context of recent results and ISW science, and conclude.

Throughout this work, we shall consider a flat $\Lambda$CDM scenario compatible with {\it Planck} DR3 cosmology: $\Omega_{\mathrm{b}}=0.049117$, $\Omega_{\Lambda}=0.684857$, $\Omega_{\mathrm{m}}=0.315143$, $n_{\mathrm{s}}=0.963$ and $h=0.6726$. Unless otherwise specified, we assume galaxy bias equal to unity ($b_g=1$). We use Greek indices running from 0 to 3 to denote spacetime variables, and Latin indices running from 1 to 3 to refer to the spatial part of a four-tensor.

\section{Post-Newtonian description of Angular Redshift Fluctuations}

\label{sec:PostNewt}

\subsection{The ARF observable}

In this Section, we briefly introduce the reader to the ARF first presented in Letter I. In that work, the ARF field was defined as 
\beq \label{eq:ARF_defobsI}
    \bar{z}+(\delta z)^{\rm I} (\vnh)=\frac{\sum_{j\in \vnh} z_jW_j}{\sum_{j\in \vnh} W_j },
\eeq
where the superscript ${\rm I}$ stands for this first ARF definition, $\bar{z}$ denotes the average (monopole) redshift over the footprint, and $W_j$ corresponds to a Gaussian weight given by \\
$W_j\equiv \exp{\left[-(z_{\rm obs}-z_j)^2/(2\sigma_z^2)\right]}$. The sum index $j$ runs through all galaxies falling in a sky region pointing to $\hat{\mathbf{n}}$. The weight $W_j$ measures the distance of the redshift of the $j$-th galaxy to a Gaussian redshift shell of center $z_{\rm obs}$ and width $\sigma_z$\footnote{Like in all previous work, throughout this work we shall be referring to Gaussian windows exclusively; this is done for the sake of simplicity. We stress however that ARF can naturally be defined under any redshift window as long as it has ``ends'', i.e. it is confined to a redshift interval.}. Both $z_{\rm obs}$ and $\sigma_z$ are chosen conveniently by the observer, and motivate the tomographic character of the ARF. 
\\\\
However, the denominator of Eq.~\ref{eq:ARF_defobsI} can be noisy (and even zero) for surveys with sparse sampling, and for practical reasons a second definition was proposed in Letter II:
\beq
\label{eq:ARF_defobsII}
(\delta z)^{\rm II} (\vnh)=\frac{\sum_{j\in \vnh} (z_j-\bar{z})W_j}{\langle \sum_{j\in \vnh} W_j \rangle_{\vnh}},
\eeq
where the ensemble average $\langle ... \rangle_{\vnh}$ takes place over all pixels $\vnh$ in the survey's footprint, i.e., it is an area average. It can be shown that both definitions yield the same expressions under linear theory of cosmological perturbations, although their sensitivity to potential systematics biasing the observed number of tracers is not the same: it can be easily seen that $(\delta z)^{\rm I}$ is robust against multiplicative systematics, unlike $(\delta z)^{\rm II}$, while both of them are robust against additive systematics that do not appreciably vary under the redshift shell. For the sake of simplicity, we shall hereafter adopt the second definition for the ARF, Eq.~\ref{eq:ARF_defobsII}. 

\subsection{Post-Newtonian theoretical derivation}


In the post-Newtonian limit, we can define the observed redshift of the $j$-th observed galaxy as 
\begin{equation}
    z_j = z_H + (1+z_H) \mathbf{v} \cdot \mathbf{\hat{n}},
    \label{eq:zobs_N}
\end{equation}  where $z_H$ stands for the Hubble drift redshift $1 + z_H = 1/a$, $a$ is the expansion factor, and $\mathbf{v}$ denotes the (physical/proper) peculiar velocity vector in units of the speed of light. Applying either of the two definitions of the ARF above, we obtain

\[
\bar{z}+\delta z(\hat{\mathbf{n}})=\mathcal{F}\left[z_{H}\right]+\mathcal{F}\left[b_{g} \delta_{\mathrm{m}}\left(z_{H}-\mathcal{F}\left[z_{H}\right]\right)\right]+
\]
\begin{equation}
\phantom{xxxxxxxx}
\mathcal{F}\left[\left(z_{\phi}+\mathbf{v} \cdot \hat{\mathbf{n}}\left(1+z_{H}\right)\right)\left(1-\frac{d \log W}{d z}\left(z_{H}-\mathcal{F}\left[z_{H}\right]\right)\right)\right] + \mathcal{O}\left(2^{\mathrm{nd}}\right),
\end{equation}
where $b_g$ is the linear galaxy bias and $\delta_m$ is the matter density contrast. Note that we use linear perturbation theory to derive the previous previous equation, i.e. it only includes terms at first order in $\delta_{\rm m}$ and $\vv$. Here we have defined the normalised functional
\begin{equation}
\label{eq:functional1}
\mathcal{F}[Y]=\frac{\int \mathrm{d} r r^{2} \bar{n}(r) W\left(z_{obs}-z_H[r]\right) Y(r)}{\int \mathrm{d} r r^{2} \bar{n}(r) W\left(z_{cen}-z_H[r]\right)}=\frac{1}{\bar{n}_{ang}} \int \mathrm{d} r\,r^{2} \bar{n}(r) W\left(z_{cen}-z_H[r]\right) Y(r),
\end{equation}
with $\bar{n}_{ang}$ the average angular number density of galaxies\footnote{Henceforth we shall refer to ``galaxies'' to any type of extragalactic luminous matter tracer, be it galaxies or quasars.} under the considered Gaussian shell $W(z)$ centered upon $z_{\rm cen}$, and $r$ the comoving radial distance. Note here $z_\phi$ refers to redshift fluctuations of gravitational origin\footnote{Negligible in this approach, we shall compute their amplitude explicitly in Sect.~\ref{sec:results}.}.

\section{General Relativistic derivation of the ARF}
\label{sec:ARFinGR}

In this section, we derive the linear order, general relativistic corrections to ARF in a flat universe, and we compute their sensitivity to different aspects of cosmological physics.

\subsection{The framework for source number counts}

Our starting point will be the observed number counts of galaxies per unit redshift and solid angle, given by $n(\vnh,z)dzd\Omega$, to which we shall refer hereafter as ``angular density fluctuations" or ADF. As mentioned above, there is abundant literature computing the GR corrections to this observable \cite{Yoo08,Yoo09,Yoo10,BonvinDurrer11}, although in this work we shall follow closely the approach of CL11. We next briefly revisit the main findings of CL11 that we need to build upon in order to compute the GR corrections of ARF, and defer the reader to Appendix~\ref{app:ADF} for a more detailed derivation. 

CL11 work in the Newtonian gauge in a flat Friedmann-Lemaître-Robertson-Walker (FLRW) Universe described by the metric
\begin{equation}
d s^{2}=g_{\mu\nu} dx^{\mu}dx^{\nu}=a^{2}(\eta)\left\{-(1+2 \psi) \mathrm{d}\eta^{2}+(1-2 \phi) \delta_{i j}d x^{i} d x^{j}\right\},
\end{equation}
where $a(\eta)$ is the cosmological scale factor, $\psi$ and $\phi$ are the scalar potentials perturbing this otherwise homogeneous metric, and where vector and tensor perturbations are ignored. CL11 compute the linear-order estimation of the redshift measured by an observer from a source emitting light at conformal time $\eta$:
\begin{equation}
\label{eq:redshift_M}
    \begin{split}
     1+z(\eta)&=\left(\frac{a_{o}}{a(\eta)}\right)\left\{1+\left[v_{i} e^{i}-\psi\right]_{o}^{g}+ \int_{\eta_0}^{\eta} \mathrm{d} r\left[(\dot{\psi}+\dot{\phi})\right]\right\} \\
     &=\frac{a_{o}}{a(\eta)}\left(1+\psi_{o}-\psi+\hat{\mathbf{n}} \cdot\left[\mathbf{v}-\mathbf{v}_{o}\right]+\int_{\eta_{o}}^{\eta}(\dot{\phi}+\dot{\psi}) \mathrm{d} \eta^{\prime}\right),
    \end{split}
\end{equation}
where dot variables refer to derivatives with respect to $\eta$, and the vector $\mathbf{v}$ is the spatial vector of the four-velocity of a comoving observer (with the ``$o$'' subscript denoting the observer's position in space-time).
CL11 map {\em observed} redshifts (which are dependent upon metric perturbations) into {\em perturbed} radial coordinates of the sources. In this way, setting $\eta=\eta_s+\delta \eta$ for a source at observed redshift $z_s$, such that $1+z_s=a_0/a(\eta_s)$, the perturbation to the conformal time ($\delta \eta$) assigned to that source reads
\beq
\label{eq:pert_eta}
\mathcal{H}(\eta_s)\delta \eta \equiv \Delta z(\eta_s)=\psi_{o}-\psi+\hat{\mathbf{n}} \cdot\left[\mathbf{v}-\mathbf{v}_{o}\right]+\int_{\eta_{o}}^{\eta_s}(\dot{\phi}+\dot{\psi}) \mathrm{d}  \eta^{\prime} +\mathcal{H}_{0} \delta \eta_{0},
\eeq
and this can be translated into the radial, comoving distance assigned to that source as
\beq
\label{eq:r_pert}
r(\mathbf{\hat{n}}, z_s) = r_s + \delta r = \eta_o - \eta_s - \delta \eta - \int_{\eta_{o}}^{\eta_{s}}(\phi+\psi) \mathrm{d} \eta^{\prime}.
\eeq

With this in mind, CL11 computed the angular power spectra associated to ADF as an integral of the curvature power spectrum $\mathcal{P}(k)$ and the squared ADF transfer function,
\beq 
\label{eq:power_M}
C_{\ell}^{\rm ADF} = \frac{2}{\pi}\int dk\,k^2\mathcal{P}(k)|\Delta^{{\rm ADF},\, W}_{\ell}(k)|^2,
\eeq
where the ADF transfer function is given by 
\beq \label{eq:number_M}
\begin{split} 
    &\Delta_{N, l}^{{\rm ADF},\,W}(k)=\int_{0}^{\eta_{o}} \mathrm{~d} \eta\left[W(\eta)\left(\delta_{N} j_{\ell}(k r)+\frac{k v}{\mathcal{H}} j_{\ell}^{\prime \prime}(k r)\right)+W_{\delta \eta}(\eta)\left[\psi j_{\ell}(k r)+v j_{\ell}^{\prime}(k r)\right]\right. \\
    &\left.+(\dot{\psi}+\dot{\phi}) j_{\ell}(k r) \int_{0}^{\eta} W_{\delta \eta}\left(\eta^{\prime}\right) \mathrm{d} \eta^{\prime}+(\phi+\psi) j_{\ell}(k r)\left(\int_{0}^{\eta}(2-5 s) \frac{W\left(\eta^{\prime}\right)}{r^{\prime}} \mathrm{d} \eta^{\prime}\right.\right.\\
    &\left.\left.+\frac{l(l+1)}{2} \int_{0}^{\eta} \frac{r^{\prime}-r}{r r^{\prime}}(2-5 s) W\left(\eta^{\prime}\right) \mathrm{d} \eta^{\prime}\right)+W(\eta) j_{\ell}(k r)\left(\frac{1}{\mathcal{H}} \dot{\phi}+\psi+(5 s-2) \phi\right)\right].
\end{split}
\eeq
This line-of-sight integral is conducted under the window function $W(\eta)=W(z)(1+z)\mathcal{H}$, with $W(z)$ the observed redshift window function of the sources, and contains the metric scalar potentials $\psi$ and $\phi$, among other quantities (like the parameter lensing magnification bias $s$) that are defined in CL11 and in Appendix~\ref{app:ADF}. In order to compute the ARF angular power spectra, and/or the cross-correlation of the ARF with any other cosmological field (like, e.g., the CMB temperature anisotropy field), we must compute the ARF transfer function, for which Eq.~\ref{eq:number_M} is our starting point. 

\subsection{Modifications required by the ARF}
\label{sec:ARF_mod}

For ARF the sources of the anisotropy are not built upon the number of matter probes (galaxies) only, but upon number-weighted redshift fluctuations with respect to an average redshift ($\bar{z}$) computed under a redshift shell that, for simplicity, we take to be symmetric and Gaussian (Letter I). We thus modify the redshift window function present in Eq.~\ref{eq:number_M} as 
\beq
\mathcal{W}(z,\bar{z}) \equiv W(z)(z-\bar{z}),
\label{eq:newWz}
\eeq
with $\bar{z} = \int dz\, W(z)\,z$ for a normalized source window function $W(z)$. 

In practice, when implementing these modifications in the code {\tt CAMB sources}, we had to further modify the integrals associated to the spherical Bessel function derivatives ($j_{\ell}^\prime(kr)$, $j_{\ell}^{\prime\prime}(kr)$): these derivatives do not appear in the code since they are avoided via integration by parts, and this latter step had to be repeated under the new window function $\mathcal{W}(z,\bar{z})$. Since the code is written in the so-called {\it CDM gauge} (or {\it zero acceleration frame}), we had to transform the equations to this frame, for which use of the {\tt sympy}-based {\tt symbolic} module of the python wrapper of {\tt CAMB sources} was made. A detailed description of these changes can be found in Appendix~\ref{ap:B}.

\section{Results} \label{sec:results}

\begin{figure}[tbp]
\centering
\includegraphics[width=1. \textwidth]{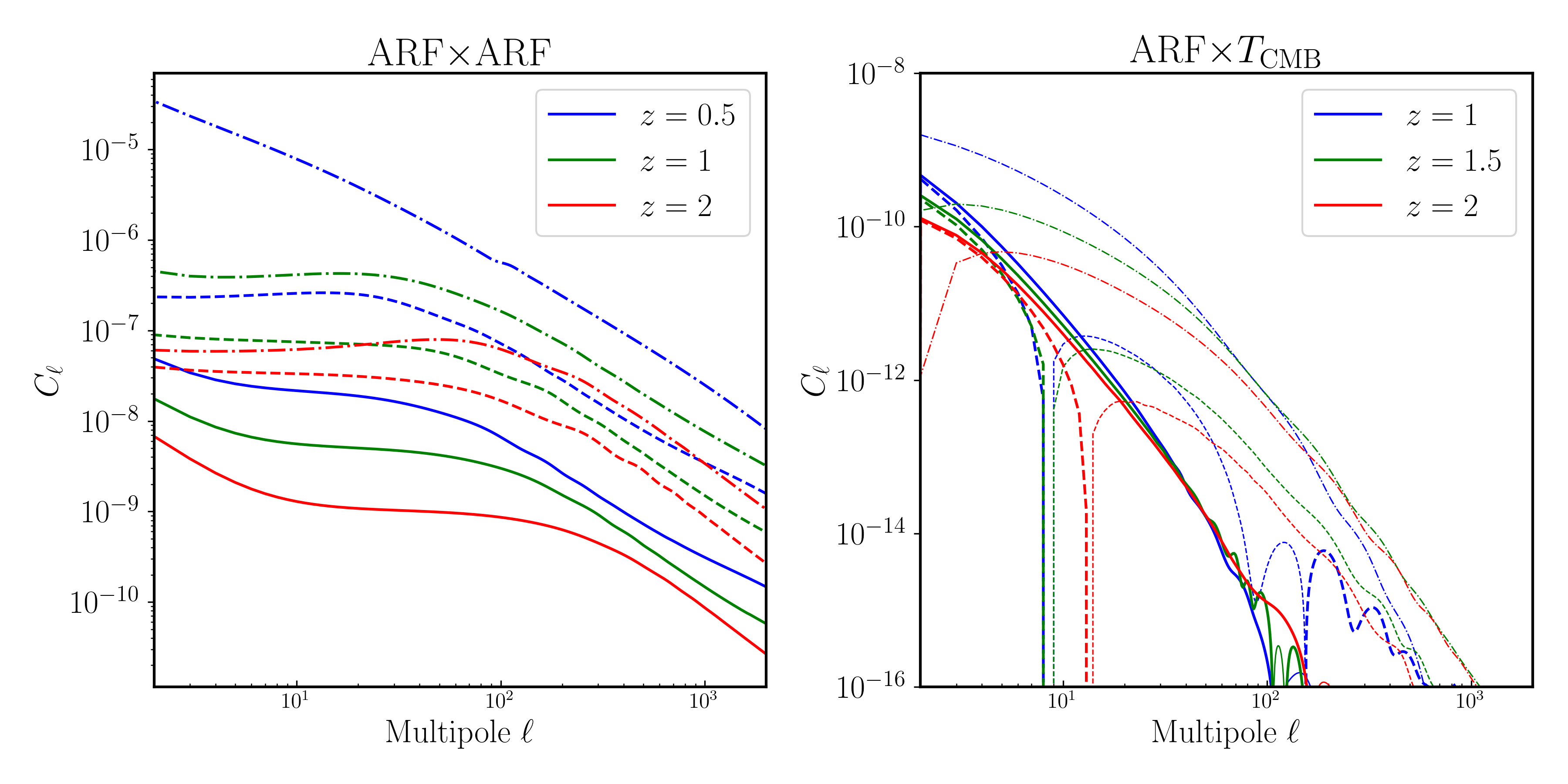}
\caption{\textit{Left panel:} Angular power spectrum for ARF for a Gaussian window function at different redshifts with widths $\sigma_z=0.01$ (solid), $\sigma_z=0.1$ (dashed) and $\sigma_z=0.35$ (point-dashed). \textit{Right panel:}  Cross-correlation power spectrum of ARF with CMB temperature for windows with widths $\sigma_z=0.25$ (solid), $\sigma_z=0.5$ (dashed) and $\sigma_z=0.8$ (point-dashed). Both cases assume $s=0$. Thin and thick lines display negative and positive results, respectively.}.
\label{fig:auto_cross}
\end{figure}

In Fig.~\ref{fig:auto_cross} we display illustrative examples of the ARF auto angular power spectra and the ARF cross-power spectra with primordial CMB fluctuations for redshift windows with different central redshifts and widths. We are adopting $s=0$ in this case. For the auto ARF power spectra (left panel), we find very similar shapes to those found in Letter I, with a flat profile at low multipoles and a decay on smaller angular scales (high multipoles). Only for the intermediate width ($\sigma_z=0.1$) we find some sensitivity of the ARF to the baryonic acoustic oscillations (BAO), since for this width the ARF are sensitive to the radial gradient of source density on scales close to the BAO scale. The same argument applies for the turn-over peak of the matter power spectrum $P_{m}(k)$, which becomes visible only for redshift widths close to the linear scale ($\sim 1/k_{\rm peak}$) at which $P_m(k)$ shows its maximum ($\sigma_z=0.35$, or $1/k\sim 1\,h(z)$~$h^{-1}$~Gpc, with $h(z)$ the dimensionless Hubble parameter $h(z)\equiv H(z)/H_0$).

The ARF cross correlation with the CMB intensity/temperature is shown in the right panel of Fig.~\ref{fig:auto_cross}, and depending on the shell width and central redshift may flip sign at different angular scales. For low central redshifts of the shells the ARF$\times$CMB cross-correlation is small and un-measurable. Only at high redshifts ($z>1.5$) and large widths ($\sigma_z>0.3$) an {\em anti-}correlation between the ARF and the CMB becomes measurable. As it will be shown below in Sect.~\ref{sec:cmbX}, this anti-correlation (which is enhanced by the lensing term) is built on top of the ISW effect in the CMB field, since it vanishes when the ISW term is dropped from the CMB anisotropy computation. The ARF provide thus an alternative window to dark energy via an ISW anti-correlation, although this requires probing the density field at high redshifts ($1<z<3$).


\begin{figure}[tbp]
\centering
\includegraphics[width= 1.1\textwidth]{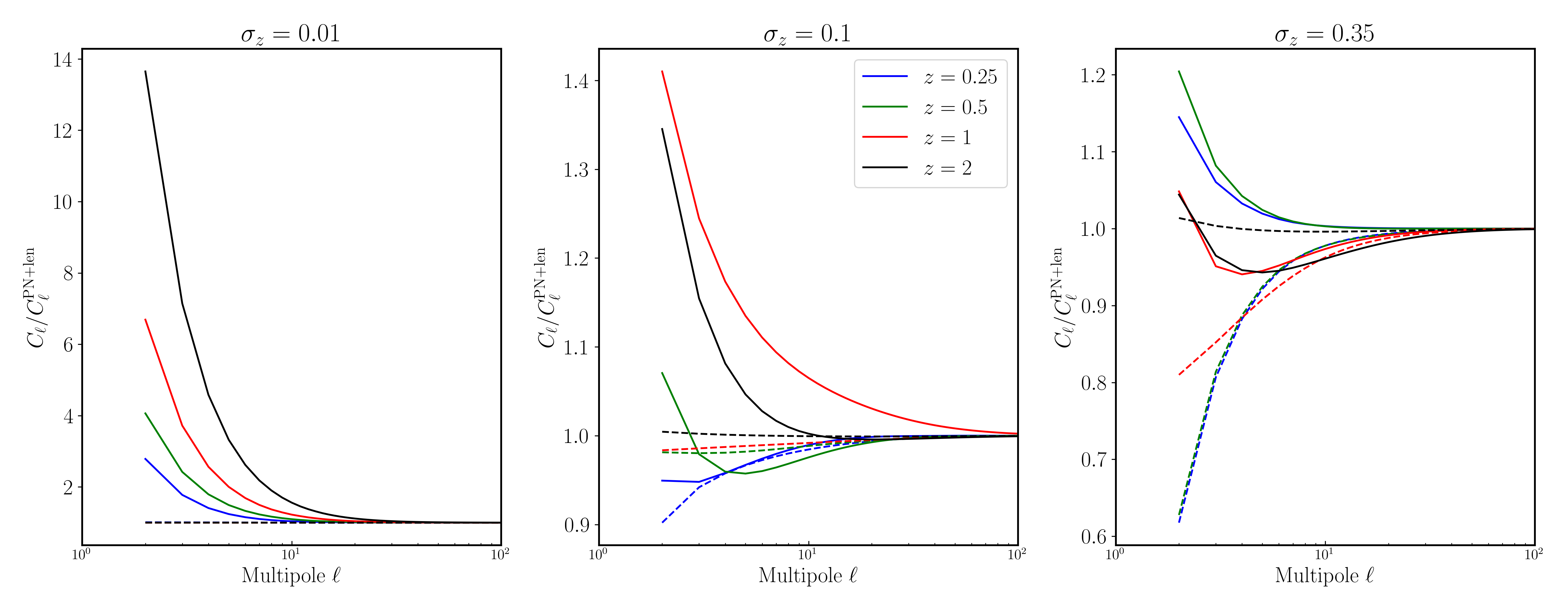}
\caption{Relative importance of GR corrections on ARF (solid) and ADF (dashed) power spectra at different redshifts for $s=0$. Each panel display the result for different Gaussian widths. }
\label{fig:amplitudes}
\end{figure}
\begin{figure}[tbp]
\centering
\includegraphics[width=1.05 \textwidth, height= 13cm]{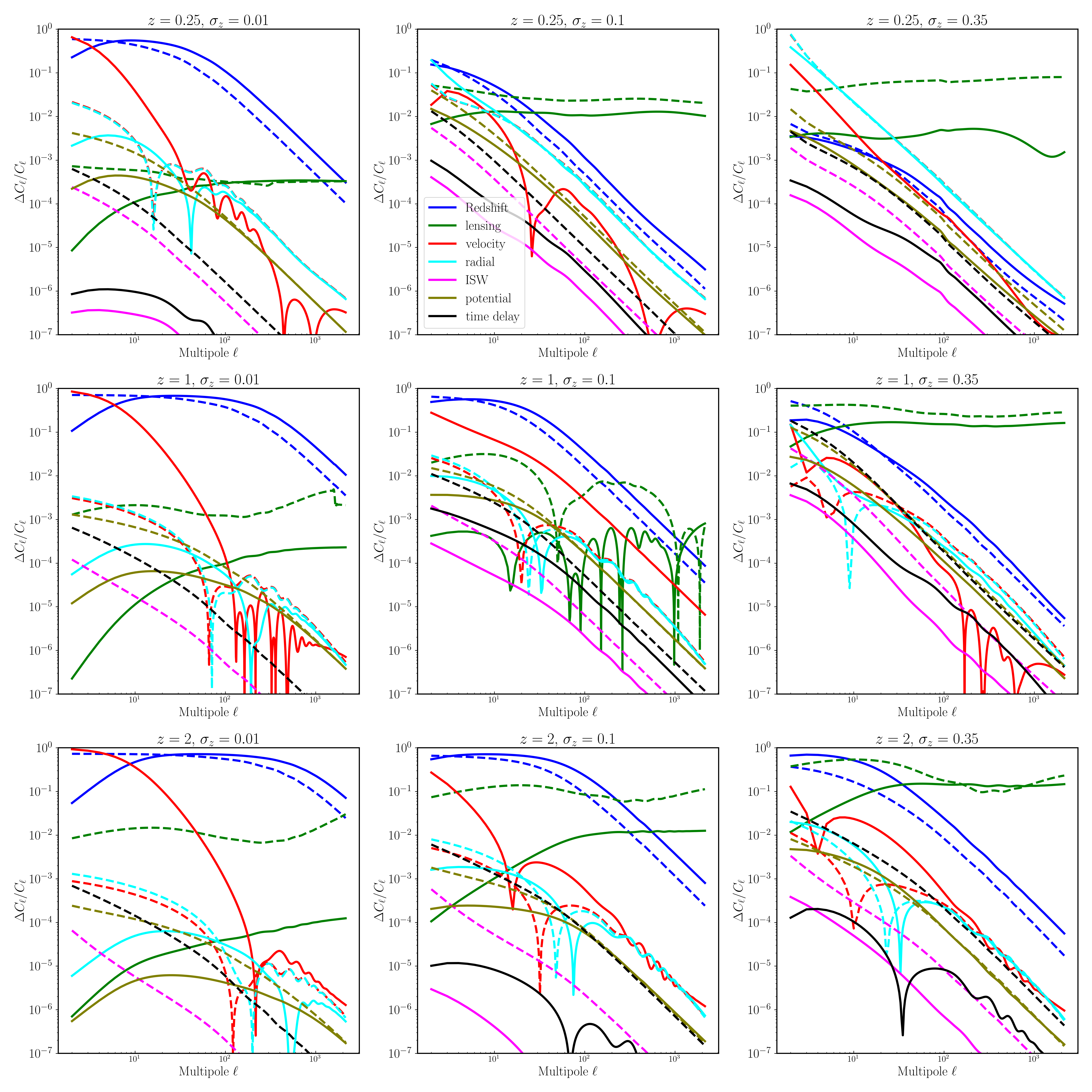}
\caption{Fractional difference compared to the full result for the ARF (solid lines) and ADF (dashed lines) angular power spectra for $z=0.25, \, 1$ and $2$. Here we show the relative amplitudes of the different terms in Eq.~\ref{eq:number} when neglecting all other contributions: \textit{RSD} refers to Redshift Space Distortions (radial derivative of velocity, quoted as the {\it redshift} term in Appendix~\ref{ap:B}), \textit{lensing} to the convergence term, \textit{velocity} to the terms proportional to $\mathbf{v \cdot \hat{n}}$ without the radial displacement (included in \textit{radial}), and we break the terms of gravitational origin into the \textit{ISW} term, the (Shapiro) \textit{time delay} term, and \textit{potential} term for all remaining contributions from potentials. We also consider $s=0$.}
\label{fig:fractional1}
\end{figure}

\subsection{ARF angular auto-spectra}

The ratio of the angular power spectrum with all relativistic corrections ($C_{\ell}^C$) and considering only the {\it redshift} and {\it lensing} contributions\footnote{See Appendix~\ref{ap:B} for further details on these terms.}  ($C_{\ell}^{PN+len}$), as it is usually considered in the literature, is displayed in Fig.~\ref{fig:amplitudes}. Here we show results for ADF and ARF in dashed and solid lines respectively. It shows the importance of other corrections, which we will see are velocity-related corrections, at low multipoles for both ADF and ARF. We can see however that such GR corrections are more important for ARF than for ADF in most cases, although restricted to very large angular scales ($\ell\lesssim 10$). As we move to higher $\ell$, the lensing terms start to dominate over the other corrections both for ADF and ARF.

For a deeper insight into this behaviour, we study the different contributions to the linear-order ARF/ADF angular power spectrum separately. In Fig.~\ref{fig:fractional1} we single-out the different terms which appear in Eq.~\ref{eq:number}, and show the fractional error compared to the full result $\Delta C_\ell / C_\ell^C$ for the ARF (solid) and ADF (dashed) angular power spectra. Here, $\Delta C_\ell \equiv \abs{C_\ell^{\,\,(i)}-C_\ell^{NC}}$ refers to the absolute difference of the power spectra obtained after adding only the correction of the $i$th term minus the un-corrected case ($C_\ell^{NC}$). 

The {\it redshift} term refers to the radial derivative of the peculiar velocity ($\partial \mathbf{v}/\partial r$), and is usually included in a post-Newtonian approach (like, e.g., the one in Letter I). This term has a major contribution for both ADF and ARF. Interestingly, ARF seem to be significantly more sensitive than ADF to all other velocity-related corrections (namely the {\it velocity} and the {\it radial} terms) than ADF, at least for narrow shells. On the other hand, the relative importance of the {\it lensing} term (given by green curves in Fig.~\ref{fig:fractional1}) seems comparable for both ADF and ARF: for both observables the impact of lensing increases with the shell width, and becomes the most important linear correction on small angular scales/high multipoles. For wide shells, the ARF {\it lensing} term is below that of ADF at low redshifts, although both become similar at intermediate ($z=1$) and high ($z=2$) redshifts.



Finally, all terms related to gravitational potentials ({\it time delay}, {\it ISW}, and {\it potentials}) are generally smaller than the previous ones, just as CL11 found for ADF: the amplitude of these terms increases with the shell width (like for the {\it lensing} term). Note that these corrections for ARF are significantly smaller than for ADF, for reasons we discuss below.

\subsection{ARF$\times$CMB angular cross- spectra}
\label{sec:cmbX}

In this sub-section we study the cross-correlation of the ARF with the primordial CMB temperature $T$, E-mode polarization ($E$), and lensing deflection ($\phi$) maps. Let $X$ be any of the three CMB observables $T,E$, and $\phi$, and $Y$ either the ADF ($\delta_g$) or the ARF ($\delta_z$) fields. We next use the usual Fourier decomposition of an arbitrary function $f(\hat{\mathrm{n}})$ defined on the 2D-sphere, $a^{f}_{\ell,m} = \int d\hat{\mathrm{n}} Y^{\star}_{\ell,m} (\hat{\mathrm{n}})\,f(\hat{\mathrm{n}})$, where the $a^{f}_{\ell,m}$-s are the Fourier multipoles of the function $f(\hat{\mathrm{n}})$, $Y_{\ell,m} (\hat{\mathrm{n}})$ are the usual spherical harmonics, and the star ``$\star$" is denoting here ``complex conjugate". The cross power spectrum between the fields $X$ and $Y$ is then defined as $C^{X,\,Y}_{\ell} = \langle a^{X}_{\ell,m} (a^{Y}_{\ell,m})^\star\rangle$, with $\langle ... \rangle$ denoting ensemble averages, and whose exclusive dependence upon $\ell$ follows from the (assumed) statistical isotropy of the fields. For any pair of the $X$ and $Y$ fields, we define $\chi^2_{\ell}$ as
\begin{equation}
    \chi^2_{\ell} \equiv \frac{(C_{\ell}^{X,Y})^2 (2\ell+1)f_{\rm sky}}{C_\ell^{XX}C_\ell^{YY} + (C_\ell^{X,Y})^2 },
    \label{eq:chisq_ell_def}
\end{equation}
where all non-cosmological sources of uncertainty (instrumental noise or foreground signal in the CMB fields, shot noise in ADF/ARF) are being ignored, and where the factor $(2\ell+1)f_{\rm sky}$ factor denotes the numbers of degree of freedom per multipole $\ell$. In what follows we shall assume that $f_{\rm sky}=1$. This ratio $\chi^2_{\ell}$ shows, per multipole $\ell$ and for Gaussian $X$ and $Y$ fields, the squared signal-to-noise (S/N) ratio of the cross angular spectrum multipole and its uncertainty, i.e., $(C_{\ell}^{X,Y})^2/\sigma^2_{C_{\ell}^{X,Y}}$.

\begin{figure}[tbp]
\centering
\includegraphics[width= \textwidth]{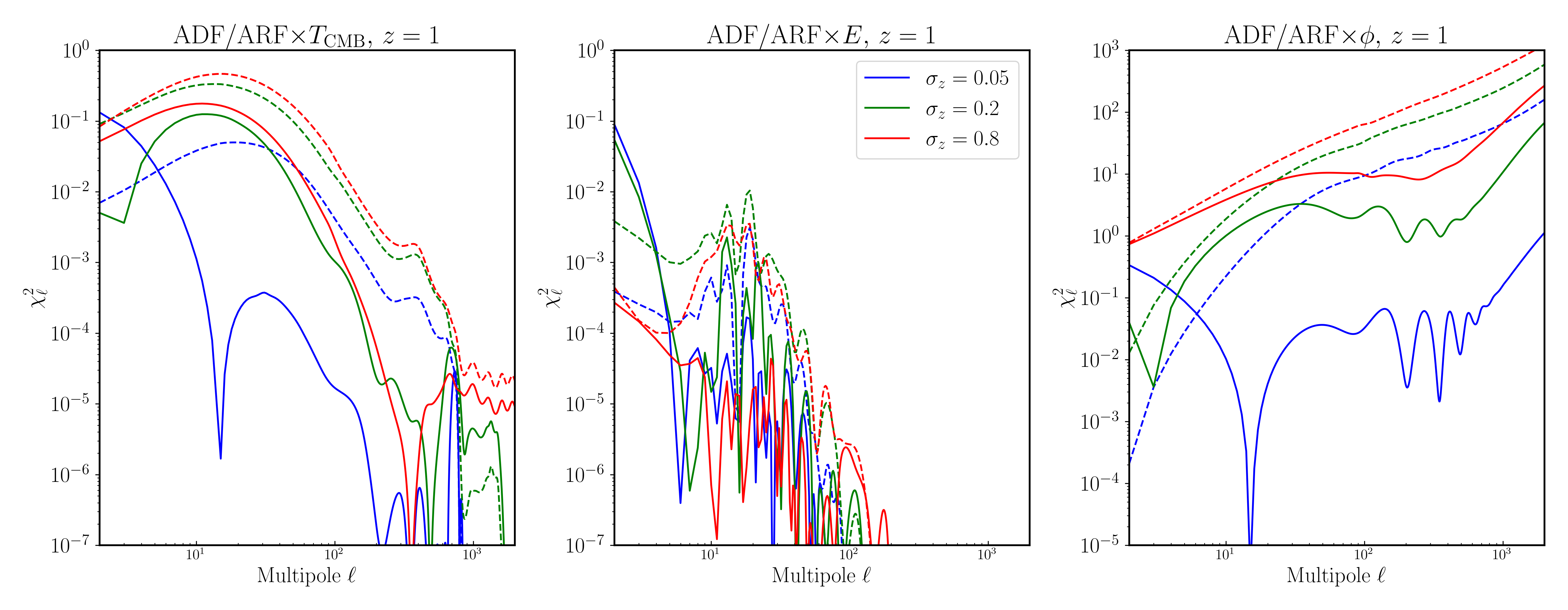}
\caption{Signal-to-noise ratio per multipole ($\chi^2_{\ell}$) of the ARF (solid) and ADF (dashed) cross-correlation with CMB temperature ($T$, left panel), $E$-mode of CMB polarization (middle panel), and lensing deflection potential $\phi$ at $z=1.0$, for $\sigma_z=0.05,$ $0.2$, and $0.8$. In this case and all subsequent cases we consider $s=0$.}
\label{fig:SN1}
\end{figure}

As shown in Fig.~\ref{fig:SN1}, the ARF display non trivial correlations with CMB observables. In particular, the left panel of this plot is showing that for large widths ($\sigma_z=0.8$), the ARF show a similar (albeit lower) level of (anti-)correlation to CMB anisotropies than the standard ADF observable. It turns out that this anti-correlation of the ARF with CMB temperature anisotropies (already shown in the right panel of Fig.~\ref{fig:auto_cross}) is triggered by the ISW. The dominant, {\it density} term of the ARF overlaps with the evolving potentials at $z\lesssim 2$. Unlike ADF, the $(z-\bar{z}) \delta_g$ term in the ARF kernel makes this observable sensitive to radial/redshift/time asymmetries under the redshift shell, and since only those potentials in the low redshift wing of the Gaussian shell ($z<\bar{z}$) are actually evolving, typically {\em negative} amplitudes of the ARF field will correlate with regions giving rise to positive ISW amplitudes, thus giving rise to this anti-correlation. The sum of Eq.~\ref{eq:chisq_ell_def} from all multipoles ($\chi^2\equiv \sum_{\ell=2}^{\ell_{\rm max}}\chi^2_{\ell}$) equals, for $z=1$ and cross-correlations with CMB temperature, $\chi^2=(0.8)^2$, $(1.8)^2$, and $(5.6)^2$ for ADF and $\sigma_z=0.05,\, 0.2,\,$ and $0.8$, respectively, while the corresponding figures for ARF and $z=2$ are $\chi^2=(0.4)^2$, $(0.5)^2$, and $(4.9)^2$. This shows that the ARF contains additional sensitivity to the ISW effect for large widths at high redshifts, as we shall explore further below.

The level of correlation of both ADF and ARF with the $E$-type of CMB polarization is rather low (see middle panel of Fig.~\ref{fig:SN1}), giving rise to negligible cumulative values of $\chi^2_{\ell}$: the sum of this statistics over all multipoles up to $\ell_{\rm max}=2\,000$ never reaches the value of unity. On the contrary, the ARF are highly correlated with the lensing potential $\phi$, although typically at not such a high level as the ADF. When integrating this linear-theory predictions up to $\ell_{\rm max}=2\,000$, the cumulative S/N for the ADF/ARF$\times \phi$ cross-correlations reaches or exceeds the level of a hundred for $\sigma_z\geq 0.1$.

\begin{figure}[tbp]
\centering
\includegraphics[width=10cm]{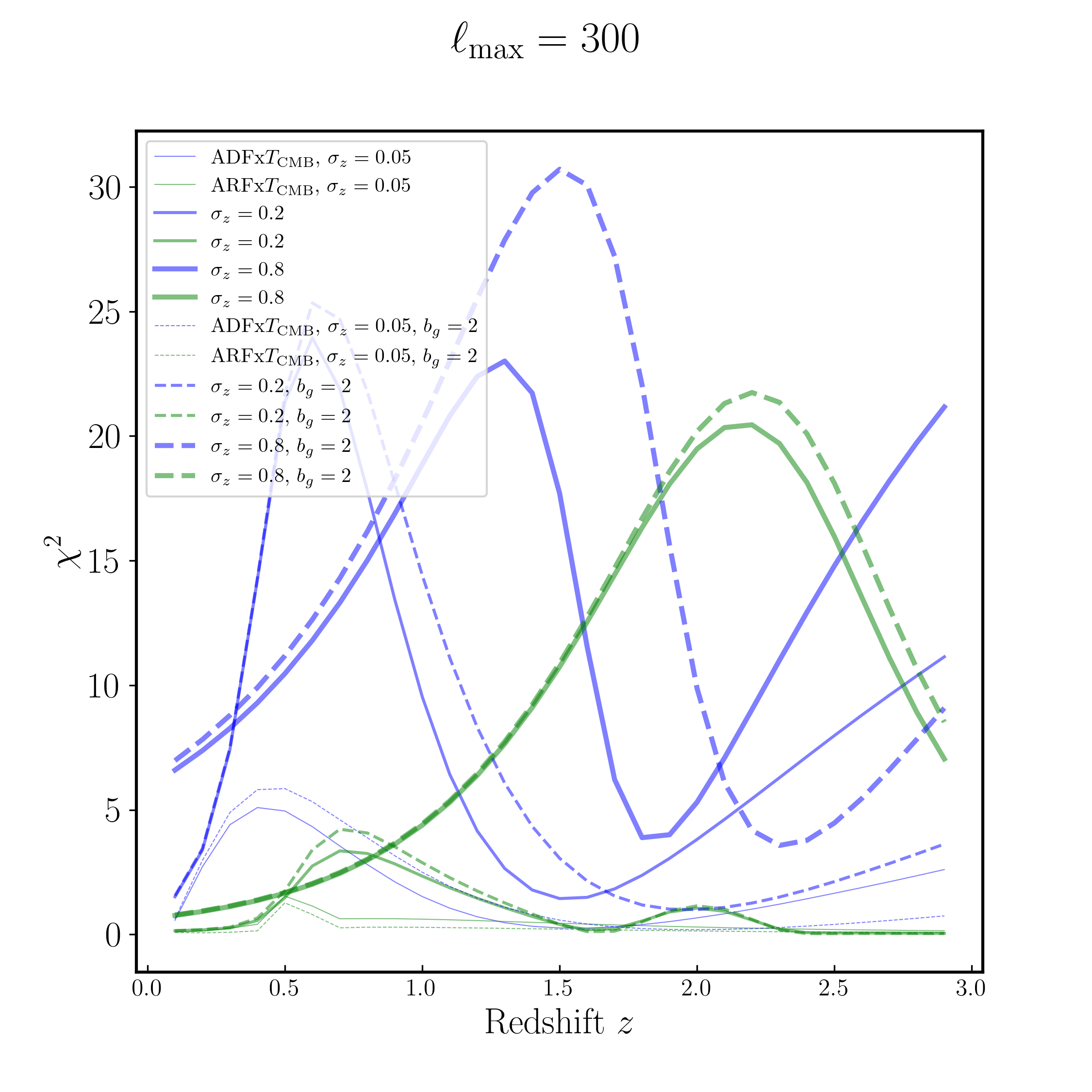}
\caption{Cumulative $\chi^2$ for the ADF/ARF$\times T$ angular cross-correlation when considering one single shell versus its central redshift. We are considering different shell widths, which scale with the thickness of the lines. Blue (green) colours refer to ADF (ARF), and dashed lines consider a linear, constant bias of $b_g=2$.  All multipoles from $\ell=2$ up to $\ell=300$ have been included when computing $\chi^2=\sum_{\ell_{\rm min}}^{\ell_{\rm max}} \chi^2_{\ell}$. We have verified that these cross-correlation amplitudes are built upon the ISW effect, since this $\chi^2$ statistic falls at the level of unity when switching off the ISW contribution in our modified version of the {\tt CAMB sources} Boltzmann code. This also motivates the choice for $\ell_{\rm max}=300$, since most of the ISW signal is contained at low multipoles ($\ell<100$).  }
\label{fig:SNoneshell}
\end{figure}

\subsubsection{A new window to the ISW effect}

Given the sensitivity of the ARF to the ISW on the large angular scales, we conduct here a more detailed study on how ISW constraints can be improved by adding ARF in cross-correlation studies. 

In Fig.~\ref{fig:SNoneshell} we consider one single redshift shell of varying widths ($\sigma_z=0.05, 0.2,$ and $0.8$) and placed at the central redshifts given by the abscissa axis. We also study the impact of a galaxy bias by including a constant, linear bias value of $b_g=2$ (depicted by dashed lines). Since we are looking at the ISW-induced ADF/ARF cross correlation with CMB temperature anisotropies, we limit the cumulative sum of $\chi^2_{\ell}$ up to $\ell_{\rm max}=300$. This plot is showing a different but complementary sensitivity of ARF compared to ADF: the latter show larger levels of ISW-induced correlation at intermediate to low redshifts ($z\in[0.5,1.5]$), from intermediate width to thick shells ($\sigma_z=0.2,\,0.8$). ARF instead show significant ($\chi^2\sim 20$) angular ISW-induced (anti-)correlations at higher redshifts (typically $z\simeq 2$), and only for larger widths ($\sigma_z=0.8$). At very high redshifts ($z>2.5$), when the $\chi^2$ for ARF and $\sigma_z=0.8$ starts to fall, corresponding statistic for ADF grows again: we have checked this is due to an {\em anti-}correlation between the ADF and the ISW anisotropies, extending up to high redshifts that cannot be easily accessed by upcoming LSS surveys due to the impact of reasonable levels of shot noise and values of the magnification bias parameter $s$. For the ARF$\times T$ anti-correlation under a shell centred upon $z=2$ and $\sigma_z=0.8$, the shot-noise of survey of average galaxy density of $\bar{n}_{\rm ang}\sim 10^{7}$~sr$^{-1}$ (or $\sim 3000$ galaxies per square degree) should degrade the ideal $\chi^2$ from $\sim 5.1^2$ down to $\sim 4.3^2$ ($b_g=1$). In Table~\ref{tab:tab1} we show how the ISW S/N degrades in ADF/ARF$\times T$ cross-correlation analyses for different levels of shot noise and luminosity function slope $s$. However, detailed predictions for a given LSS survey should include precise estimations from both the number density and the bias versus redshift. We can also see from Fig.~\ref{fig:SNoneshell} that the bias impacts more strongly the ADF$\times T$ than the ARF$\times T$ cross-correlations $\chi^2$ statistics: this is due to the different scaling of the ADF/ARF auto- and cross-angular power spectra when introducing a bias greater than one. Both terms increase, but to a different extent that depends on the cross talk between the {\it density} and {\it lensing} terms, which are dominant at these relatively large widths. 

\begin{table}
\centering
\begin{tabular}{|c|c|ccccc|}
\hline
\hline
Probe & $\bar{n}_{\rm ang}$ [sr]$^{-1}$ & & &  S/N ($\ell=300$) & &  \\
\hline
 & & $s=-0.8$ & $s=-0.4$ & $s=0$ & $s=0.4$ & $s=0.8$ \\
\hline
    & $10^6$ & $3.6$ & $0.8$ &  $2.3$ & $0.8$ & $3.6$ \\
ADF & $10^7$ & $5.8$ & $5.4$ & $4.1$ & $2.2$ & $6.1$ \\
    & $10^8$ & $5.9$ & $5.6$ &$4.6$ & $3.3$ & $6.8$ \\
\hline
    & $10^6$ & $1.0$ & $1.6$ & $2.2$ & $1.6$ &  $1.0$ \\
ARF & $10^7$ & $5.2$ & $4.8$ &  $4.3$ &  $3.4$ &  $2.3$ \\
    & $10^8$ & $5.7$ & $5.4$ & $5.0$  & $4.3$ & $3.0$ \\
\hline
\hline
\end{tabular}

\caption{\label{tab:tab1} Degradation of the S/N for ADF- \& ARF$\times$ ISW anti cross-correlation for different levels of shot-noise. We approximate the shot-noise angular power spectra as a constant given by $C_{\ell}^{\rm SN}=1/\bar{n}_{\rm ang}$ and $C_{\ell}^{\rm SN}=\sigma_z^2/\bar{n}_{\rm ang}$ for ADF and ARF, respectively. For both ADF/ARF observables, we adopt a single Gaussian shell of width $\sigma_z=0.8$, and whose central redshifts are located at $z=3$ and $z=2$ for ADF and ARF, respectively. These are roughly the central redshifts where the corresponding $\chi^2$ show a local maximum for each probe. We are considering different values for the magnification bias parameter $s$, including negative ones (see discussion).}

\end{table}

\begin{figure}[tbp]
\centering
\includegraphics[width= \textwidth]{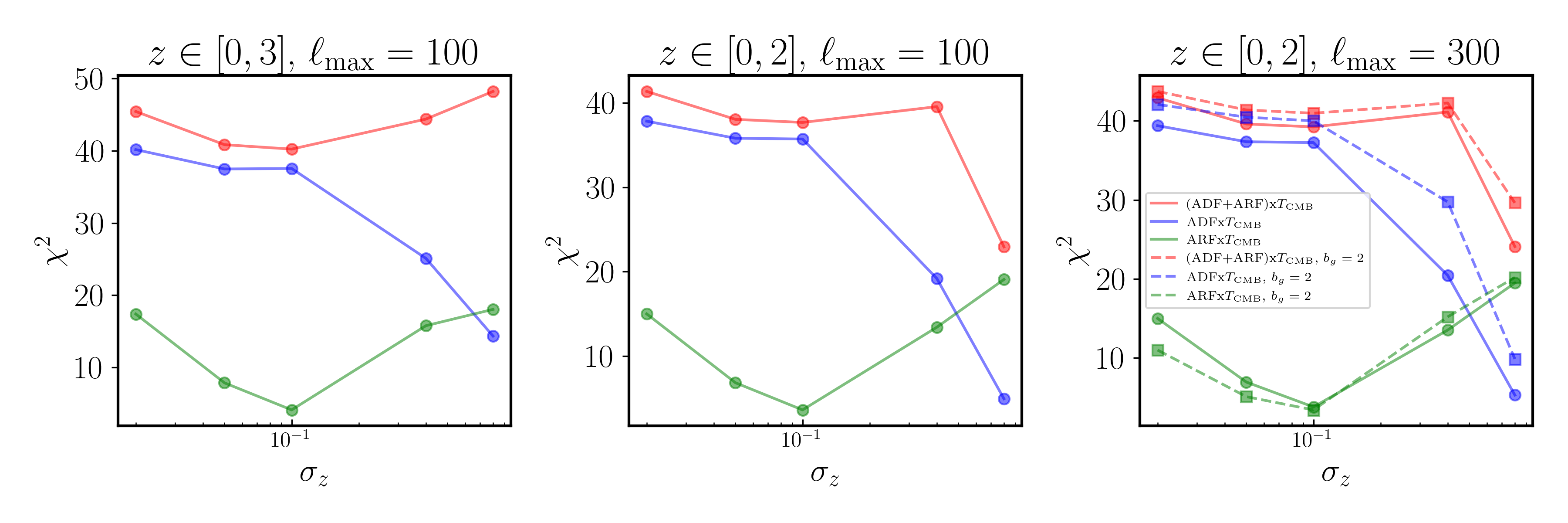}
\caption{Cumulative $\chi^2$ statistics (defined in Eq.\ref{eq:chisqMshell}) accounting for the cross-correlation of ADF, ARF, and ADF+ARF observables with CMB intensity anisotropies ($T$) for different configurations of redshift Gaussian shells. All redshift shells are taken to have the same width $\sigma_z$ given in the $X$-axis, and sweep the redshift range given in the titles at the panels' top with a regular shell separation of $3\sigma_z$. Blue, green, and red curves and symbols refer to $\mathbf{d}=\mathbf{C}_{\ell}^{\mathrm{ADF},T}$, $\mathbf{C}_{\ell}^{\mathrm{ARF},T}$, or $(\mathbf{C}_{\ell}^{\mathrm{ADF},T},\mathbf{C}_{\ell}^{\mathrm{ARF},T})$, respectively. The differences between the middle and left panel show the (small [ARF] to moderate [ADF]) impact of decreasing $z_{\rm max}$ from $3$ to $2$ on the wide shell configurations. The right panel considers the case of $\ell_{\rm max}=100$, leaving previous results virtually unchanged, and also studies the inclusion of a galaxy bias $b_g=2$ (dashed lines). In all cases we adopt $s=0$. }
\label{fig:allshellsISW}
\end{figure}

We next address the question on how much sensitivity on the ISW can be gained by adding the ARF to the ADF in a cross-correlation analysis. We consider a data vector $\mathbf{d}$ containing the cross angular power spectra of the type $\mathbf{C}^{\mathrm{ADF},T}_{\ell}$, $\mathbf{C}^{\mathrm{ARF},T}_{\ell}$, or $(\mathbf{C}^{\mathrm{ADF},T}_{\ell}, \mathbf{C}^{\mathrm{ARF},T}_{\ell})$, where the boldface indicate vectors containing all multipoles from $\ell_{\rm min}=2$ up to $\ell_{\rm max}$, for {\em all} redshift shells under consideration. The Gaussian shell configuration is such that the shell centers sample a redshift interval $[z_{\rm min},z_{\rm max}]$, where all equally-spaced Gaussian redshift shells have widths given by the abscissas axis of Fig.~\ref{fig:allshellsISW}. Wide shells with centers close to $z_{\rm max}$ will incorporate galaxies at redshifts above $z_{\rm max}$. We adopt $3\,\sigma_z$ for the separation between adjacent redshift shells, so the number of shells is  given by $N_{\rm shells}=(z_{\rm max}-z_{\rm min})/(3\sigma_z)$. For our choices of $\sigma_z$, this means that the number of shells range from 50 down to one. After neglecting all non-cosmological sources of uncertainty (shot noise, CMB foreground emission, survey systematics, etc) we compute the covariance matrix $\mathbf{\cal{C}}\equiv \langle \mathbf{d} \mathbf{d}^t \rangle - \langle \mathbf{d} \rangle \langle \mathbf{d}^t \rangle $ (here $\mathbf{d}^t$ denotes the ``transposed" version of the $\mathbf{d}$ array). Our covariance matrix thus accounts for the non-zero correlation of ADF, ARF in all redshift shells under study. This way we define the $\chi^2$ statistic as 
\begin{equation}
    \chi^2 \equiv  \mathbf{d}\, \mathbf{\mathcal{C}^{-1}} \mathbf{d}^t.
    \label{eq:chisqMshell}
\end{equation}

In the left panel of Fig.~\ref{fig:allshellsISW} we take $\ell_{\rm max}=100$ in our cross-correlation analysis, and adopt $z_{\rm min}=0$, $z_{\rm max}=3$. The estimates values of $\chi^2$ for ADF are displayed by the blue line and blue circles: this line display a rather flat behaviour for low values of $\sigma_z$ ($\sigma_z=0.02$--$0.1$, $N_{\rm shells}=12$--$50$), but decreases significantly when considering broader redshift shells (down to $\chi^2\simeq 12$ for $\sigma_z=0.8$). Interestingly, the same statistics for the ARF$\times T$ cross-correlation shows a distinct pattern: as the green circles show, it has a minimum around $\sigma_z=0.1$, but then increases up to $\chi^2\simeq 16$ for $\sigma_z=0.8$. This agrees with our previous result in Fig.~\ref{fig:SNoneshell} pointing a higher ARF sensitivity to the ISW for wide shells at $z\simeq 2$--$3$. When we combine both observables (ADF and ARF) in this cross-correlation analysis, we obtain the $\chi^2$ statistics given by the red line: for a thin-shell configuration the typical gain of adding the ARF is about a $\simeq 10$~\% in $\chi^2$, so a $\sim 5$~\% in S/N,  although this improvement increases to $\sim 150$~\% in $\chi^2$ for wide redshift shells ($\sigma_z=0.8$).

Provided that many LSS surveys may not be able to sample the $z\sim 3$ universe densely, in the middle panel of Fig.~\ref{fig:allshellsISW} we decrease $z_{\rm max}$ to $2$, finding that the main pattern found in the left panel is not significantly affected: the $\chi^2$ for the ADF$\times T$  correlation decreases slightly (particularly when losing the wide shells at $z>2$), but the one for ARF$\times T$ remains roughly unchanged. The $\chi^2$ for the joint ADF+ARF analysis reflects the changes on the ADF side. Finally, in the right panel of Fig.~\ref{fig:allshellsISW} we increase $\ell_{\rm max}$ to $300$, finding no visible differences with the previous choice of $\ell_{\rm max}=100$ (as expected for ISW cross-correlations, \cite{hernandez-monteagudo2008_ImplementationFourierMatched}). In this panel we again estimate the impact of galaxy bias by adopting $b_g=2$ (dashed lines). As in Fig.~\ref{fig:SNoneshell}, we find that $b_g$ has a stronger impact in ADF than on ARF.
Its impact in the joint ADF+ARF $\chi^2$ statistics is rather modest (around 5~\%). 

Figs.~\ref{fig:SNoneshell},\ref{fig:allshellsISW} show that the ARF provide an different view to the ISW, alternative to that provided by ADF: by studying ARF on wide shells at $z\sim 2$ we should be able to constrain ISW with with about half the accuracy (in units of $\chi^2$)
achieved by cross-correlation analyses with ADF/2D clustering at $z\lesssim 1$.

\begin{figure}[tbp]
\centering
\includegraphics[width= \textwidth]{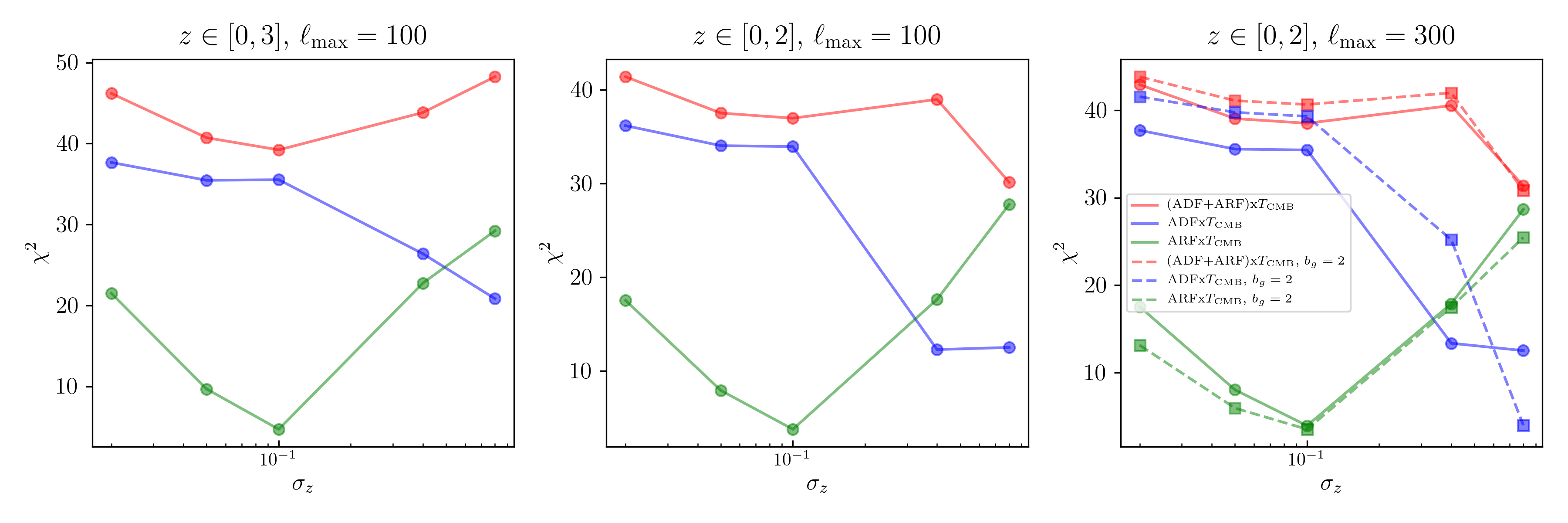}
\caption{Same as in Fig.~\ref{fig:allshellsISW}, but for $s=-0.2$. See the discussion section for the possibility of having negative values of $s$ in samples defined under colour cuts.  }
\label{fig:allshellsISW_sm0p2}
\end{figure}


\section{Discussion and Conclusions}
\label{sec:discussion}

Our analyses have shown that relativistic corrections for ARF are particularly relevant for all terms involving peculiar velocities ({\it redshift, radial,} and {\it velocity}). Lensing is also important, specially for wide shells, on the small angular scales, while other related gravitational potential terms are negligible. These results match the expectations derived from the Post-Newtonian approach of the ARF already presented in Letter I: given the extra factor $(z-\bar{z})$ present in the ARF kernel relative to the ADF kernel, this observable is typically sensitive to those cosmological fields that {\em vary} significantly under the redshift shell width, thus becoming ideal for probing cosmic densities and velocities for shell widths of the order $\sigma_z\sim 0.01$--$0.1$. Gravitational potentials typically vary on yet larger scales, and would a priori require wider shells: we find that indeed the contribution from gravitational potential terms ({\it time delay}, {\it ISW}, {\it potentials}) increases under larger redshift shell widths (e.g., one can easily compare the amplitudes of these terms in the left and right columns in Fig.~\ref{fig:fractional1}), but their intrinsic low amplitude (compared to typical density and velocity contributions) makes their detection challenging. The fact that the $(z-\bar{z})$ kernel makes the ARF sensitive to only radial/redshifts gradients of the sources also reflects in the right panel of Fig.~\ref{fig:SN1}: ARF only pick up the radial gradient part of the $\phi \times \delta_g$ cross-correlation (missing the constant or redshift-symmetric part), and for this reason $\chi^2_{\ell}$ is lower for ARF$\times \phi$ than for ADF$\times \phi$.

As in Letter I, we are finding that corrections involving the velocity field are typically higher for ARF than for ADF, particularly for narrow ($\sigma_z\sim 0.01$--$0.02$) redshift shells for which opposite-sign cancellation of line-of-sight integrals are less likely. ARF constitute thus a sensitive tool to measure peculiar velocities and the nature of gravity on cosmological scales, as already demonstrated in Letter II. Our efforts will next focus on the estimation of ARF statistics in the midly non-linear regime, and for that we shall resort to cosmological numerical simulations.

During the final edition stage of this work, \cite{matthewson2022_RedshiftWeightedgalaxy} presented a parallel work where the ARF GR corrections at first order are computed in an arbitrary gauge. These authors are concerned about the gauge-dependence of redshift fluctuations as defined in Eq.~\ref{eq:redshift}. In their work they argue that redshift fluctuations must be defined with respect to an average redshift ($\bar{z}$), estimated from the observed redshifts of the galaxy sample under $W(z)$. They conclude with a definition of their ``redshift-weighted number counts" which coincides with the ARF definition provided at the beginning of Letter I.\footnote{Even if they rename this observable as ``redshift-weighted galaxy number counts" rather than ``density-weighted angular redshift fluctuations" introduced in Letter I (which thereafter we have dubbed as "ARF" for the sake of simplicity), they use {\em exactly} the same observable introduced in Letter I: the (galaxy) density-weighted redshift fluctuations with respect an average redshift ($\bar{z}$) estimated from a galaxy sample under a given redshift window function $W(z)$}. Their computation of the linear-order GR corrections coincides with ours in the formulation (the addition of the new modified function $\mathcal{W}(z)\equiv (z-\bar{z}) W(z)$), and also in the numerical results: when comparing their Fig.~1, the amplitudes of the total contribution to the angular power spectrum lies in absolute agreement with ours. We also find perfect agreement if we identify their ``large scale gravitational potential terms" with the joint contribution of our {\it time delay}, {\it ISW}, and {\it potential} corrections. Likewise, their RSD angular power spectrum coincides with the addition of our {\it redshift}, {\it radial}, and {\it velocity} terms. Finally, after adopting the value $s=0.81$ (motivated by their Eq.~4.2 for $s(z)$ evaluated at $z\sim 1$), we obtain a lensing contribution to the $C_{\ell}$-s whose amplitude and shape is practically indistinguishable from theirs. 

A relevant aspect of our analysis involves the ARF$\times$ISW cross correlation. The ISW constitutes one of the very few windows we have towards the nature of Dark Energy, and a very distinct one, since it is the only observable measuring the impact of the universe's expansion rate on the evolution of gravitational potentials. However, it involves so large scales that it is heavily affected by cosmic variance, and so far the statistical significance for the ISW is relatively weak ($\sim 4$ if one includes the CMB lensing map, $\sim 3$ only if one restricts the cross-correlation analysis to the CMB temperature map and LSS surveys, \cite{collaboration2016_Planck2015results}). This type of analyses also involves the largest scales observable in the current universe, where GR effects should be more important but whose sampling is most difficult.  
 In this context, we have shown that ARF's sensitivity to the ISW complements nicely that of the traditional approaches based on ADF: ARF must be strongly {\em anti-}correlated to CMB temperature anisotropies when probing the LSS at $z\sim 2$ under wide redshift shells, and this is a significantly different universe to that where ADF are more sensitive to the ISW ($z\in[0.5,1]$). An interesting aspect lies in the dependence of the statistical significance of the ARF$\times$ISW cross-correlation on the source's luminosity function slope $s$. A priori, one should always expect $s$ to remain positive for a flux/magnitude limited sample, and this is indeed the case in the forecasts for flux-limited surveys like Euclid and SKA (\cite{Euclid_phot_lensing, SKA_Bonaldi}). However, the value of $s$ can be negative for galaxy samples selected according to other criteria; for instance, the value of $s$ is well below zero for BOSS galaxies (see \cite{BOSS_LF_2016}). This is because the BOSS survey aimed to observe the reddest, brightest galaxies, and at fixed redshift the fraction of red galaxies decreases with magnitude. This suggests the possibility of exploring, within large photometric samples, new sub-sample definitions for which $s(z)$ becomes negative, hence enhancing the S/N of the ARF$\times$ISW cross-correlation. 
\\\

In this work we have explored the ARF in the context of general relativistic linear order of perturbations. We have followed the approach of CL11, and generalized their expressions for the source counts in the Newtonian gauge to the ARF, by including the observed redshift in the total observable under study. We have computed the linear-order general relativistic corrections, and found that all those terms relate to velocity introduce measurable effects. The corrections due to lensing are also detectable, and are actually dominant on the small angular scales for intermediate-width to wide ($\sigma_z\gsim 0.1$) redshift shells. All other terms related to the gravitational potentials are introducing negligibly small corrections in all cases. 

We have also studied the ARF cross-correlation to CMB intensity, polarization, and lensing potencial fluctuations. Similarly to ADF, the ARF show negligible correlation to the $E$-type of CMB polarization, but a high level of correlation with the $\phi$ CMB lensing potential field. This correlation for ARF$\times \phi$ is not as high as for ADF$\times \phi$ since the ARF are only picking up the time/radial gradient of the density--potential correlation due to the $z-\bar{z}$ part of the kernel.  
The ARF present alternative correlation properties with CMB temperature fluctuations in the presence of the ISW effect. The {\it density} term in the ARF kernel gives rise to an anti-correlation with the ISW temperature anisotropies if the redshift shells are wide ($\sigma_z \gsim 0.5$) and are placed at a high redshift ($z\sim 2$). The ideal S/N of this anti-correlation lies at the $4$--$5$ level, comparable (albeit lower) to that of the ADF$\times T$ cross-correlation, although it arises at complementary redshift and shell width ranges. By combining ADF and ARF in cross-correlation studies with the CMB intensity maps, the $\chi^2$ statistic against the null hypothesis (no ISW) is increased by about $\sim 5$--$10$~\% when conducting LSS tomography with using narrow shells in $z\in [0,2]$, and by $\sim 150$~\% when resorting to a few, wide ($\sigma_z\sim 0.8$) redshift shells in the same redshift range. The ARF thus provide a novel, potentially powerful, and complementary tool to test the nature of dark energy in the late universe.


\acknowledgments

We thank Dr.~Juan Betancourt, Dr.~Evencio Mediavilla, and Dr.~Jordi Cepa for the useful and insightful comments that they provided as members of the committee for the under-graduate master thesis that gave rise to this work. We also thank Prof.~Roy Marteens, Dr.~Daniele Bertacca, and Dr.~Chema Diego for useful discussions. CHM acknowledges partial support from the Spanish Ministry of Science, Innovation and Universities (MCIU/AEI/FEDER, UE) through the grant PGC2018-097585-B-C21. JCM likewise acknowledges partial support from the partner grant PGC2018-097585-B-C22.

\bibliographystyle{mnras}
\bibliography{biblio} 


\appendix

\section{General relativistic derivation of the source number counts}
\label{app:ADF}

This section outlines the main findings of CL11 on the GR first order corrections to source number counts (or ADF), and are included here as a quick reference for the main text. 
\\\\
We shall consider a flat Friedmann-Lemaître-Robertson-Walker (FLRW) Universe described by the metric
\begin{equation}
d s^{2}=g_{\mu\nu} dx^{\mu}dx^{\nu}=a^{2}(\eta)\left\{(1+2 \psi) \mathrm{d}\eta^{2}-(1-2 \phi) \delta_{i j}d x^{i} d x^{j}\right\}.
\end{equation}
written under the conformal-Newtonian gauge\footnote{The choice of gauge coincides with that of CL11 in their work (although it differs from the gauge used in the code {\tt CAMB sources}).}, with $a$ the expansion parameter at conformal time $\eta$ and the two scalar potentials $\phi$ and $\psi$
\cite{mukhanov1992_TheoryCosmologicalperturbations}.

\subsection{Redshift perturbations}
We must express $n(\vnh,z)\,(z-\bar{z})\,dzd\Omega$ in term of covariant quantities which transform under coordinate transformation as dictated by the space-time metric. The observed redshift can be trivially expressed as a function of the contraction of covariant 4-vectors following its definition
\begin{equation}
1+z_{s}=\frac{\left(k^{\mu} u_{\mu}\right)_{s}}{\left(k^{\mu} u_{\mu}\right)_{o}},
\label{eq:reds_def1}
\end{equation}
where $k^{\mu}$ is the photon null momentum and $u^{\mu}$ denotes the 4-velocity of galaxies, and subscripts $s$ and $o$ refer to quantities evaluated at the source's and the observer's positions, respectively. 
\\
For a photon moving along a geodesic $x^{\mu}(\lambda)$ with $\lambda$ an \textit{affine parameter} along the geodesic, we can define its null-momentum $k^{\mu}=\frac{d x^{\mu}}{d \lambda}$ as
\begin{equation}
\left\{\begin{array}{l}
    k^{0}=-\frac{\bar{\nu}}{a}(1+\delta \nu) \\ k^{i}=\frac{\bar{\nu}}{a}\left(e^{i}+\delta e^{i}\right),
\end{array} \right.
\end{equation}
in the observer's rest frame\footnote{All subsequent calculation will be performed in the observer's rest frame.}. Here we define $\bar{\nu}$ as the photon frequency\footnote{In this case, the bar denotes a \textit{background} quantity and not a mean.} and $e^{i}$ as the photon propagation direction measured by the observer in a homogeneous universe ($\vec{e} \equiv \hat{\mathbf{n}}$); and $\delta \nu$, $\delta e^{i}$ its respective dimensionless corrections as we expand the null vector to first order in perturbations, which can be related to the metric perturbations $\phi,\psi$ by solving the null and geodesic equations.
\\
For this choice of coordinates we can define the 4-velocity of a comoving observer as
\begin{equation}
u^{\mu}=\frac{d x^{\mu}}{\sqrt{-d s^{2}}}=\frac{d x^{\mu}}{d \tau},
\end{equation}
with components $u^0=a^{-1}(1-\psi)$ and $u^{i}=a^{-1}v^{i}$. Here $\tau$ the proper time along the observer's worldline and $v^{i}(\eta,\hat{\mathbf{n}})\ll 1$ the physical, peculiar velocity in units of the speed of light. As previously stated the redshift at conformal time $\eta$ along the line of sight will be given by the ratio of photon energies at the source's and observer's positions (Eq.~\ref{eq:reds_def1}), so up to linear order
\begin{equation}
\label{eq:redshift}
    \begin{split}
     1+z(\eta)&=\left(\frac{a_{o}}{a(\eta)}\right)\left\{1+\left[v_{i} e^{i}-\psi\right]_{o}+ \int_{\eta_0}^{\eta} \mathrm{d} r\left[(\dot{\psi}+\dot{\phi})\right]\right\} \\
     &=\frac{a_{o}}{a(\eta)}\left(1+\psi_{o}-\psi+\hat{\mathbf{n}} \cdot\left[\mathbf{v}-\mathbf{v}_{o}\right]+\int_{\eta_{o}}^{\eta}(\dot{\phi}+\dot{\psi}) \mathrm{d} \eta^{\prime}\right),
    \end{split}
\end{equation}
where dot variables refer to derivatives with respect to the conformal time $\eta$. Furthermore, in a homogeneous and isotropic universe, at the observer's position $a(\eta_o)=a_o$, which is a constant. However, taking into account perturbations in the conformal time at the observer's position due to local gravitational potential effects
\beq \label{eq:timepert}
a_o=a(\eta_o+\delta\eta_o)\approx a(\eta_o)+ \dot{a_o}\delta\eta_o=a(\eta_o)(1+\mathcal{H}_{o} \delta \eta_{o}),
\eeq
 with $\mathcal{H}_{o}=\dot{a}(\eta_o)/a(\eta_o)$ the conformal Hubble parameter\footnote{The addition of this constant term only impacts the monopole and is thus unobservable.}. This way we can express the redshift to linear order as



\beq \label{eq:red_pert}
1+z(\eta)=\frac{a_o}{a(\eta)}(1+\Delta z),
\eeq
with $\Delta z$ depending on the gravitational potentials and peculiar velocities at observer's and source's positions and on the evolution of these potentials.\footnote{Although we have taken the spatial part of the metric to be $g_{ij}=a^2(\eta)(1-2\phi)\delta_{ij}$, the result yields for a more general choice $g_{ij}=a^2(\eta)(1-2\phi)\bar{g}_{i j}$, where $\bar{g}_{i j}$ is defined as the 3-space background metric} 
Following again CL11, we can map these perturbations of the observed redshift to perturbations of the conformal time we assign to sources. Setting $\eta = \eta_s + \delta \eta$ for a source at observed redshift $z_s$, such as $1+z_s = a_o/a(\eta_s)$, following Eq. (\ref{eq:red_pert}) one can express, to linear order,
\beq
\mathcal{H}(\eta_s)\delta \eta = \Delta z(\eta_s)=\psi_{o}-\psi+\hat{\mathbf{n}} \cdot\left[\mathbf{v}-\mathbf{v}_{o}\right]+\int_{\eta_{o}}^{\eta_s}(\dot{\phi}+\dot{\psi}) \mathrm{d}  \eta^{\prime} +\mathcal{H}_{0} \delta \eta_{0},
\eeq
and in the same way CL11 write the (perturbed) radial position of a photon at $z_s$ as
\beq
r(\mathbf{\hat{n}}, z_s) = r_s + \delta r = \eta_o - \eta_s - \delta \eta - \int_{\eta_{o}}^{\eta_{s}}(\phi+\psi) \mathrm{d} \eta^{\prime}.
\eeq

Note that in this expression the lensing-induced transversal shift in the apparent position of a source does not appear, since it does not impact its radial coordinate. We conclude this subsection by stressing that redshift perturbations present in the {\em observed} redshift are mapped into the (perturbed) radial coordinates assigned to observed sources.

\subsection{Corrections to the angular source number counts}

We again follow the approach of CL11 for expressing the galaxy angular number counts $n(\hat{\mathbf{n}}, z)$ in a covariant form. Hereafter we shall refer to these standard 2D, angular source number counts anisotropies as ``angular redshift fluctuations" or ADF.
The determinant of Jacobi map $\det \mathcal{D}$ \cite{sachs1961_GravitationalWavesGeneral, schneider1992_GravitationalLenses, lewis2006_WeakGravitationallensing} is used to relate the area covered by the unit of solid angle at the observer's and at the source's position: a ray bundle within solid angle $d\Omega_o$ at the observer's position projects an invariant area $d\Omega_o \det \mathcal{D}_o$. Thus the volume element sampled by the ray bundle at the source's position equals $d\Omega_o \det \mathcal{D}_o k_{\mu}u^{\mu}_s d\lambda$ when the affine parameter changes by an amount $d\lambda$, making the wavefront advance by an amount of $k_{\mu}u^{\mu}_s$. The quantity $J^{\mu}=n_s u^{\mu}_s$ expresses the source 4-current, with $n_s$ the source proper number density as seen in the source rest frame given by $u^{\mu}_s$. CL11 next write the amount of galaxies swept in the volume element at the source's position as
\beq
dN(\hat{\mathbf{n}}, z) = d\Omega_o n(\hat{\mathbf{n}}, z) \biggl|\frac{dz}{d\lambda}\biggr|d\lambda  = d\Omega_o \det \mathcal{D}_o n_s k_{\mu}u^{\mu}_s d\lambda,
\eeq
leading to the general result
\beq
n(\hat{\mathbf{n}}, z)=\operatorname{det} \mathcal{D}_{o} k_{a} J^{a}\left|\frac{\mathrm{d} \lambda}{\mathrm{d} z}\right|.
\eeq
CL11 compute the linear order perturbations $\delta_n(\hat{\mathbf{n}},z,m < m_{\star})$ to the observed angular number counts $n(\hat{\mathbf{n}},z,m < m_{\star})=\bar{n}(z,m < m_{\star}) [1+\delta_n(\hat{\mathbf{n}},z,m < m_{\star})]$ per angle and redshift interval, providing the following expression in which terms are sorted according to their amplitude/importance: 

\begin{equation} 
\label{eq:numCount}
\begin{split}
    &\delta_n(\hat{\mathbf{n}}, z, m<m_{\star})=\delta_{N}\left(L>\bar{L}_{\star}\right)-\frac{1}{\mathcal{H}} \hat{\mathbf{n}} \cdot \frac{\partial \mathbf{v}}{\partial r}+(5 s-2)\left[\kappa-\frac{1}{r} \int^{\eta_{o}}(\phi+\psi) \mathrm{d} \eta\right] \\
    &+\left[\frac{2-5 s}{\mathcal{H} r}+5 s-\frac{\partial \ln \left[a^{3} \bar{N}\left(L>\bar{L}_{\star}\right)\right]}{\mathcal{H} \partial \eta}+\frac{\dot{\mathcal{H}}}{\mathcal{H}^{2}}\right]\left[\psi+\int^{\eta_{o}}(\dot{\phi} + \dot{\psi}) \mathrm{d} \eta-\hat{\mathbf{n}} \cdot \mathbf{v}\right]\\
    &+\frac{1}{\mathcal{H}} \dot{\phi}+\psi+(5 s-2) \phi.
\end{split}
\end{equation}
This expression neglects uninteresting monopole and dipole terms, and refers to the relative source count perturbations for sources brighter than an apparent magnitude $m_{\star}$ (or conversely more luminous than a threshold luminosity $L_{\star}$). In the expression above $s$ is defined as $s=\partial \log_{10} \bar{N}(z,L>L_{\star})/\partial m_{\star}$, with $\bar{N}(z,L>L_{\star})$ the background physical (proper) number density of sources above the luminosity threshold $L_{\star}$. The convergence $\kappa$ appears in the determinant of the Jacobi map $\mathcal{D}$ defined above (see e.g. \cite{lewis2006_WeakGravitationallensing}), and is defined as\footnote{In the literature different definitions of the convergence has been adopted, including different terms that CL11 included separately. See \cite{yoo2009_NewPerspectivegalaxy, bernardeau2010_FullskyLensingshear}.}
\begin{equation}
    \label{eq:kappa_def}
    \kappa(\hat{\mathbf{n}},\eta)=-\frac{1}{2}\nabla^2_{\hat{\mathbf{n}}} \int_{\eta_o}^{\eta} d\eta'\frac{\eta'-\eta}{(\eta_o-\eta)(\eta_o-\eta')} (\psi + \phi),
\end{equation}
where $\nabla^2_{\hat{\mathbf{n}}}$ is the Laplacian on the 2D sphere, and $\eta_A$ refers to the conformal time at the observer's position. In Eq.~\ref{eq:numCount}, the first term ($\delta_N$) constitutes the intrinsic, Newtonian-gauge source number perturbation, which is followed by the so-called redshift space distortion (RSD) term expressing the radial gradient of the source's peculiar velocity ($\mathcal{H}^{-1}\hat{\mathbf{n}}\cdot \partial \mathbf{v} / \partial r$). This term is followed by the convergence ($\kappa$) lensing term, plus other (usually sub-dominant) terms associated to the Shapiro time delay, counts number evolution, the integrated Sachs-Wolfe effect (ISW), and other potential terms. 
\\\\
\subsection{Transfer function for ADF}
In order to compute the angular power spectrum of the source counts relative fluctuations, CL11 integrate Eq.~\ref{eq:number} along the line of sight for every mode in Fourier space, and then project the result of that integral in $k$-space against the primordial curvature power spectrum:
\beq 
\label{eq:power}
C_{\ell}^{\rm ADF} = \frac{2}{\pi}\int dk\,k^2\mathcal{P}(k)|\Delta^{{\rm ADF},\, W}_{\ell}(k)|^2.
\eeq
In this equation, $\mathcal{P}(k)$ is the primordial curvature power spectrum, and $\Delta^{{\rm ADF},\, W}_{\ell}(k)$ is the transfer function containing the line-of-sight integral of Eq.~\ref{eq:number} after expanding the plane waves in spherical Bessel functions ($j_{\ell}(x)$s): 
\beq \label{eq:number}
\begin{split} 
    &\Delta_{N, l}^{{\rm ADF},\,W}(k)=\int_{0}^{\eta_{o}} \mathrm{~d} \eta\left[W(\eta)\left(\delta_{N} j_{\ell}(k r)+\frac{k v}{\mathcal{H}} j_{\ell}^{\prime \prime}(k r)\right)+W_{\delta \eta}(\eta)\left[\psi j_{\ell}(k r)+v j_{\ell}^{\prime}(k r)\right]+(\dot{\psi}+\dot{\phi}) j_{\ell}(k r) \right. \\
    &\int_{0}^{\eta} W_{\delta \eta}\left(\eta^{\prime}\right) \mathrm{d} \eta^{\prime}+(\phi+\psi) j_{\ell}(k r)\left(\int_{0}^{\eta}(2-5 s) \frac{W\left(\eta^{\prime}\right)}{r^{\prime}} \mathrm{d} \eta^{\prime}+\frac{l(l+1)}{2} \int_{0}^{\eta} \frac{r^{\prime}-r}{r r^{\prime}}(2-5 s) W\left(\eta^{\prime}\right) \mathrm{d} \eta^{\prime}\right) \\
    &\left.+W(\eta) j_{\ell}(k r)\left(\frac{1}{\mathcal{H}} \dot{\phi}+\psi+(5 s-2) \phi\right)\right].
\end{split}
\eeq
Here $W(\eta)=W(z)(1+z)\mathcal{H}$, with $W(z)$ the observed redshift window function of the sources, whereas this other window function 
\beq
  W_{\delta \eta}(\eta) \equiv\left[\frac{2-5 s}{\mathcal{H} r}+5 s-\frac{\partial \ln \left[a^{3} \bar{N}\left(L>\bar{L}_{s *}\right)\right]}{\mathcal{H} \partial \eta}+\frac{\dot{\mathcal{H}}}{\mathcal{H}^{2}}\right]_{\eta} W(\eta)  
  \label{eq:wingtau}
\eeq
contains the term accounting for the source evolution versus redshift ($\propto \partial (a^3\bar{N})/\partial \eta$). In Eq.~\ref{eq:number}, $\mu\equiv \hat{\mathbf{n}}\cdot \hat{\mathbf{k}}$ terms have been integrated by parts (giving rise to derivatives of the spherical Bessel functions), and the order of double line-of-sight integrals have been switched). \\

\section{Transfer functions in the CDM frame}\label{ap:B}

Defining the Newtonian velocity as a gauge-invariant quantity $v_N=v+\sigma$ (from which one can infer that $\sigma=0$ in the Newtonian gauge), and accounting for the fact that in the zero acceleration frame the CDM velocity $v=0$, one finds that $v_N\equiv \sigma$ in the CDM frame. Moreover, in this frame $\psi=\phi$. Hence, we can rewrite Eq.~(\ref{eq:numCount}) in this frame and explicitly integrate by parts to eliminate the Bessel spherical function derivatives. As it is done in the \texttt{CAMB sources} code, we will present this result distinguishing among the different terms included in Eq.~(\ref{eq:numCount}), describing the exact changes introduced towards the computation of the ARF angular power spectrum. Among all relativistic corrections, we neglect the one associated to the evolution of sources, since we assume that the effective source redshift window function is actually a sub-sample of the total, underlying galaxy sample with a Gaussian shape. 
\\\\

{\large\textit{Density term}}\\\\
This term refers to the intrinsic, underlying perturbation in the background number density of sources under the Gaussian shell with luminosity exceeding $L_{\star}$. Following its definition in the Newtonian gauge in CL11,
\beq
\delta_{n}=b \delta_{m}^{\text {syn }}+\frac{\dot{\bar{n}}_{s}}{\bar{n}_{s}} \frac{v}{k} =
b\delta_m^{\text{syn}} + \biggl[ \frac{\mathrm{d}\ln{(a^3\bar{n_s})}}{\mathrm{d}\eta}-3\mathcal{H}\biggr]\frac{v}{k},
\label{eq:Delta_den}
\eeq
it is readily expressed in the CDM frame just after substituting $v$ with $\sigma$, as the bias (as it is classically defined) is defined in an orthogonal, synchronous and comoving ($v/k=0$) gauge \cite{wands2009_ScaledependentBiasprimordial}. After multiplication by the kernel $\left( z-\bar{z}\right)$, such that $\mathcal{W}(\eta)\equiv(z-\bar{z})W(\eta)$, we obtain the equivalent expression for the density term in the ARF transfer function:
\beq
\Delta_{\ell}^{\text{Den}}(k) = \int_0^{\eta_o} d\eta\, \mathcal{W}(\eta) \biggl\{ b\delta_m^{\text{syn}} + \biggl[ \frac{\mathrm{d}\ln{(a^3\bar{n_s})}}{\mathrm{d}\eta}-3\mathcal{H}\biggr]\frac{\sigma}{k} \biggr\} j_{\ell}(kr[\eta]).
\label{eq:Delta_rsd}
\eeq

We stress that for ARF computations we shall assume we are taking a {\em sub-sample} of the observed galaxy sample following a Gaussian redshift distribution, and thus we shall always ignore the intrinsic time evolution of the comoving density of sources ($ \mathrm{d}\ln{(a^3\bar{n}_s)}/\mathrm{d}\eta$) present in the square bracket of Eq.~\ref{eq:Delta_rsd}.
\vspace{1cm}

{\large\textit{Redshift term or redshift space distortion term}}\\\\
The term refers to the radial gradient of the line-of-sight velocity, and is expressed in Eq. (\ref{eq:number}) as the second derivative of the spherical Bessel function $j_\ell(kr)$. In order to obtain it we first note that \texttt{CAMB sources} neglects the contributions of anisitropic stress $\Pi$ in the source integration. Hence for a flat universe with zero anisotropic stress, we can express the evolution equations for the scalar shear $\sigma$ as

\beq
\dot{\sigma}=-2\mathcal{H}\sigma + \eta_k, \quad \ddot{\sigma} = (-2\dot{\mathcal{H}}+ 4\mathcal{H}^2)\sigma -2\mathcal{H}\eta_k + \frac{\hat{q}}{2},
\eeq
with $\eta_k=k\,\eta_s$ for $\eta_s$ the usual synchronous gauge scalar perturbation, and $\hat{q}=8 \pi G a^2 q$ the total heat flux so that $\dot{\eta_k}=\hat{q}/2$. We also rewrite the Friedmann equation as
\beq
\frac{\ddot{a}}{a^2}=\frac{1}{6a}\left(\hat{\rho}-3 \hat{p}\right),
\eeq
with $\hat{\rho}=8 \pi G a^2 \rho$ and $\hat{p}=8 \pi G a^2 p$ the total density and pressure parameters in a flat universe with no anisotropic stress. Having this present, CL11 code the following result for redshift term in the ADF transfer function in the CDM frame:
\beq \label{eq:redADF}
\begin{split}
    \Delta_{\ell}^{\mathrm{Redshift,\,ADF}} (k) &=  \int_0^{\eta_o} \, d\eta \frac{k v}{\mathcal{H}} j_{\ell}^{\prime \prime}(k r) W(\eta)\\
    & =\int_0^{\eta_o} d\eta \Bigg\{\Bigg[\Big(4\mathcal{H}^2 + \hat{\rho} + \frac{\hat{p}}{3}\Big)W_{\star}(\eta) + \Big(-4\dot{W_{\star}}(\eta)\mathcal{H} +\ddot{W_{\star}}(\eta)\Big)\Bigg]\frac{\sigma}{k}\\
    & \phantom{xxxx} + \Big(\frac{\hat{q}}{2}- 2\eta_k \mathcal{H}\Big) \frac{W_{\star}(\eta)}{k} + 2\dot{W_{\star}}(\eta)\frac{\eta_k}{k}\Bigg\}j_\ell(kr)\\
    & \equiv \int_0^{\eta_o} \, d\eta\, j_{\ell}(kr[\eta]) \, S^{\text{Redshift,\,ADF}} (k),
    \end{split}
\eeq
where $W_{\star}(\eta)\equiv W(\eta)/\mathcal{H}$ and the dot again denotes derivative with respect to $\eta$. Here $S_{\ell}^{\text{Redshift,\,ADF}}(k)$ the redshift source term for the ADF, and so for the ARF

\beq
\begin{split}
 \Delta_{\ell}^{\mathrm{Redshift,\,ARF}} (k) &=  
    \int_0^{\eta_o} d\eta\frac{k v}{\mathcal{H}} j_{\ell}^{\prime \prime}(k r) \mathcal{W}(\eta)\\
    & =\int_0^{\eta_o} d\eta\,j_\ell(kr[\eta])  \biggl[ \frac{1}{ka}\biggl\{\Big(5\mathcal{H}\sigma -2\eta_k \Big)W(\eta) - 2\sigma \mathcal{H}\dot{W_{\star}}(\eta)\\
    &\phantom{xxxx} + \frac{\sigma}{\mathcal{H}}\left(\frac{\hat{\rho} + 3\hat{p}}{6}\right)W(\eta)\biggr\}+ \left( z-\bar{z}\right) S_{\ell}^{\text{Redshift,\,ADF}} (k) \biggr] \\
    &\equiv\int_0^{\eta_o} \, d\eta\, j_{\ell}(kr[\eta]) \, S^{\text{Redshift,\,ARF}} (k).
\end{split}
\eeq
\vspace{1cm}

{\large\textit{Radial term}}\\\\

The so-called radial term corresponds to the $\propto v/(\mathcal{H}r)\, j_{\ell}^\prime(kr)$ term present in Eq.~\ref{eq:number}. Formally, according to this equation, this term is included in $W_{\delta\eta}\,v\,j_{\ell}^\prime(kr)$, but CL11 prefer to code it separately given its $\propto 1/r$ dependence that should make it relevant only for relatively nearby sources. For ADF, CL11 find
\beq
\begin{split}
    \Delta_{\ell}^{\mathrm{Radial,\,ADF}} (k) &=  \int_0^{\eta_o} d\eta\,W(\eta)\frac{(2-5s)\sigma}{\mathcal{H}r} j_{\ell}^\prime (kr)\\
    &= \int_0^{\eta_o} d\eta\, j_{\ell}(kr) \biggl\{ \frac{\sigma}{k}\biggl[ \frac{\dot{W_{\star}}}{r}-2\mathcal{H}\frac{W_{\star}}{r} +\frac{W_{\star}}{r^2}\biggr] +\frac{W_{\star}\eta_k}{kr} \biggr\} (2-5s)\\
    &\equiv \int_0^{\eta_o} \, d\eta\, j_{\ell}(kr[\eta]) \, S^{\text{Radial,\,ADF}} (k),
\end{split}
\eeq
For ARF, we need to introduce the modified window function $\mathcal{W}(\eta)$, yielding
\beq
\begin{split}
    \Delta_{\ell}^{\mathrm{Radial,\,ARF}} (k) &=  \int_0^{\eta_o} d\eta\,\mathcal{W}(\eta)\frac{(2-5s)\sigma}{\mathcal{H}r} j_{\ell}^\prime (kr)\\
    &= \int_0^{\eta_o} d\eta\, j_{\ell}(kr) \biggl\{ (z-\bar{z})\,S^{\text{Radial,\,ADF}}(k)-\frac{(2-5s)W_{\star}\sigma\mathcal{H}}{kra}\biggr\} \\
    & \equiv \int_0^{\eta_o} \, d\eta\, j_{\ell}(kr[\eta]) \, S^{\text{Radial,\,ARF}} (k).
\end{split}
\eeq
\vspace{1cm}

{\large\textit{The velocity term}}\\\\
This term describes the product $W_{\delta\eta}\,v\,j_{\ell}^\prime(kr)$ in the ADF case, and, according to Eq.~\ref{eq:number}, it includes the previous, radial term, which must be subtracted. CL11 write it in {\tt CAMB sources} as 
\beq
\begin{split}
    \Delta_{\ell}^{\mathrm{velocity,\,ADF}} (k) &=  \int_0^{\eta_o} d\eta\,W_{\delta\eta}(\eta)\,\sigma\,j_{\ell}^\prime (kr) -  \Delta_{\ell}^{\mathrm{Radial,\,ADF}}(k)\\
    &= \int_0^{\eta_o} d\eta\, j_{\ell}(kr) \biggl\{ \frac{\sigma}{k} \dot{W}_{\delta \eta}-2\frac{\mathcal{H}\sigma W_{\delta \eta}}{k} +\frac{W_{\delta \eta}\eta_k}{k}\biggr\} - \Delta_{\ell}^{\mathrm{Radial,\,ADF}}(k) \\
    &\equiv \int_0^{\eta_o} \, d\eta\, j_{\ell}(kr[\eta]) \, S^{\text{velocity,\,ADF}} (k).
\end{split}
\eeq
The extension for ARF trivially reads as
\beq
\begin{split}
    \Delta_{\ell}^{\mathrm{velocity,\,ARF}} (k) &=  \int_0^{\eta_o} d\eta\,\mathcal{W}_{\delta\eta}(\eta)\,\sigma\,j_{\ell}^\prime (kr) -  \Delta_{\ell}^{\mathrm{Radial,\,ARF}}(k)\\
    &= \int_0^{\eta_o} d\eta\, j_{\ell}(kr) \biggl\{ \frac{\sigma}{k} \dot{\mathcal{W}}_{\delta \eta}-2\frac{\mathcal{H}\sigma \mathcal{W}_{\delta \eta}}{k} +\frac{\mathcal{W}_{\delta \eta}\eta_k}{k}\biggr\} - \Delta_{\ell}^{\mathrm{Radial,\,ARF}}(k) \\
    &\equiv \int_0^{\eta_o} \, d\eta\, j_{\ell}(kr[\eta]) \, S^{\text{velocity,\,ARF}} (k),
\end{split}
\eeq
where $\mathcal{W}_{\delta \eta}(\eta) \equiv (z-\bar{z})W_{\delta \eta}(\eta)$.

\vspace{1cm}

{\large\textit{Time delay term}}\\\\
The extension of the time delay (or Shapiro) term to ARF from ADF is trivially found by substituting $W(\eta)$ by $\mathcal{W}(\eta)$, after noting that $\phi$ is taken equal to $\psi$ in the CDM frame (which applies to the terms following below, namely {\it lensing, ISW}, and {\it potentials}): 
\beq
\begin{split}
\Delta_{\ell}^{\text{time\,delay,\,ARF}}(k) & = \int_0^{\eta_o} d\eta\, j_{\ell}(kr[\eta]) \,2\phi\, \int_0^{\eta}d\eta'\,(2-5s)\frac{\mathcal{W}(\eta^{\prime})}{r^{\prime}}\\
 & \equiv\int_0^{\eta_o} \, d\eta\, j_{\ell}(kr[\eta]) \, S^{\text{time\,delay,\,ARF}} (k).
\end{split}
\eeq
\vspace{1cm}

{\large\textit{Lensing term}}\\\\
The lensing term contains the convergence $\kappa$ (which involves another line-of-sight integral). The counterpart for ARF is again found by introducing the modified window function $\mathcal{W}(\eta )$:

\beq
\begin{split}
\Delta_{\ell}^{\text{lensing,\,ARF}}(k) & = 
 \int_0^{\eta_o} d\eta\, j_{\ell}(kr[\eta]) \,2\phi\,
\frac{\ell (\ell+1)}{2}\int_{0}^{\eta} \mathrm{d} \eta^{\prime}  \frac{r^{\prime}-r}{r r^{\prime}}(2-5 s) \mathcal{W}\left(\eta^{\prime}\right) \\
 & \equiv\int_0^{\eta_o} \, d\eta\, j_{\ell}(kr[\eta]) \, S^{\text{lensing,\,ARF}} (k).
\end{split}
\eeq

\vspace{1cm}

{\large\textit{The ISW term}} \\\\
The ISW term accounts for time evolution of gravitational potentials \cite{sachs67} along the photon's path towards the observer. After the usual substitution of the window function, we write
\beq
\begin{split}
\Delta_{\ell}^{\text{lensing,\,ARF}}(k) & = 
 \int_{0}^{\eta_o} d\eta\,  j_{\ell}(k r[\eta])\, 2\dot{\phi}\,\int_{0}^{\eta} \mathcal{W}_{\delta \eta}(\eta^{\prime}) \mathrm{d} \eta^{\prime} \\
 & \equiv \int_0^{\eta_o} \, d\eta\, j_{\ell}(kr[\eta]) \, S^{\text{ISW,\,ARF}} (k).
\end{split}
\eeq

\vspace{1cm}

{\large\textit{The residual potentials term}} \\\\
The rest of the terms on Eq. \ref{eq:number} involving gravitational potentials are merged together. The substitution is again trivial changing the window functions and corresponding $\psi=\phi$ 
\beq
\begin{split}
\Delta_{\ell}^{\text{potentials,\,ARF}}(k) & = 
 \int_{0}^{\eta_o} d\eta\,  j_{\ell}(k r[\eta])\,  \Bigg[\mathcal{W}(\eta) \left(\frac{1}{\mathcal{H}} \dot{\phi}+\phi+(5 s-2) \phi\right) + \mathcal{W}_{\delta\eta}(\eta)\phi\Bigg] \\
 & \equiv \int_0^{\eta_o} \, d\eta\, j_{\ell}(kr[\eta]) \, S^{\text{potentials,\,ARF}} (k).
\end{split}
\eeq

\vspace{1cm}

\end{document}